%% file: main.tex
\documentclass[sigplan,10pt,nonacm]{acmart}
\setcopyright{none}
\acmConference[]{}
\acmYear{2021}

\input{config}

\usepackage{enumitem}
\usepackage{array}
\usepackage{booktabs}
\usepackage{lscape}
\usepackage{amsmath}
\usepackage{dblfloatfix} 

\newcolumntype{C}[1]{>{\centering\arraybackslash}p{#1}}
\newcolumntype{L}[1]{>{\raggedright\arraybackslash}p{#1}}
\newcolumntype{R}[1]{>{\raggedleft\arraybackslash}p{#1}}

\newcommand{\radiant}{{Radiant}\xspace}
\newcommand{\methodname}{\radiant}

\newcommand{\autonuma}{AutoNUMA\xspace}
\newcommand{\noautonuma}{AutoNUMA disabled\xspace}
\newcommand{\numa}{NUMA\xspace}
\newcommand{\pagetable}{page table\xspace}
\newcommand{\nvmm}{NVMM\xspace}

\begin{document}

\title{Page Table Management for Heterogeneous Memory Systems}

\author{Sandeep Kumar}
\affiliation{%
  \institution{Intel Labs}
  \city{Bengaluru}
  \state{India}
}
\authornote{Work done during an internship at Intel Labs, Bengaluru, India.}
\author{Aravinda Prasad}
\affiliation{%
  \institution{Intel Labs}
  \city{Bengaluru}
  \state{India}
}

\author{Smruti R. Sarangi}
\affiliation{%
  \institution{IIT Delhi}
  \city{New Delhi}
  \state{India}
}

\author{Sreenivas Subramoney}
\affiliation{%
  \institution{Intel Labs}
  \city{Bengaluru}
  \state{India}
}

\input{abstract}

\maketitle

\input{introduction}

\input{background}

\input{motivation}

\input{design}

\input{implementation}

\input{evaluation}

\input{relatedwork}

\input{conclusion}


\input{main.bbl}

\end{document}

%% file: config.tex
\usepackage{algorithm}
\usepackage{algpseudocode}
\usepackage{subcaption}
\usepackage{listings}
\usepackage{tcolorbox} 
\usepackage{dblfloatfix} 
\usepackage{xspace}
\usepackage{pifont}
\usepackage{tikz} 
\usepackage{xcolor}
\usepackage{placeins}
\usepackage{fancyvrb}
\usepackage{graphicx}
\usepackage{listings}
\usepackage{libertine}
\usepackage{zi4}

\newcommand{\one}{\ding{182}\xspace}
\newcommand{\two}{\ding{183}\xspace}
\newcommand{\three}{\ding{184}\xspace}
\newcommand{\four}{\ding{185}\xspace}

\newcommand{\memcached}{Memcached\xspace}
\newcommand{\redis}{Redis\xspace}
\newcommand{\bfs}{BFS\xspace}
\newcommand{\btree}{BTree\xspace}

\newcommand{\xsbench}{XSBench\xspace}

\lstdefinestyle{lstcode}{
language=C,
tabsize=4,
numbers=left,
showstringspaces=false,
basicstyle=\lstfont{black},
identifierstyle=\lstfont{black},
keywordstyle=\lstfont{blue},
numberstyle=\lstfont{black},
stringstyle=\lstfont{green},
commentstyle=\lstfont{brown!100},
emph={read,fsync,close,write,pread,pwrite,open,openat,lseek},
emphstyle={\lstfont{violet}},
breaklines=true,
frame=l,
captionpos=b,
belowskip=1em,
aboveskip=1em
}

\lstdefinestyle{lstcompress}{
language=C,
numberstyle=\footnotesize,
basicstyle=\ttfamily,
numbers=left,
stepnumber=1,
breaklines=true,
showstringspaces=false,
captionpos=b,
frame=l,
tabsize=4,
numbers=left,
emphstyle={\lstfont{violet}},
}

\newtcolorbox{tbox}[3][]
{
    colframe = #2!25,
    colback  = #2!10,
    coltitle = #2!20!black,  
    title    = {#3},
    #1,
}

%% file: abstract.tex
\begin{abstract}
	
Modern enterprise servers are increasingly embracing tiered memory systems with a combination of low latency DRAMs and large capacity but high latency non-volatile main memories (NVMMs) such as Intel's Optane DC PMM. Prior works have focused on efficient placement and migration of data on a tiered memory system, but have not studied the optimal placement of {\pagetable}s. 

Explicit and efficient placement of {\pagetable}s is crucial for large memory footprint applications with high TLB miss rates because they incur dramatically higher page walk latency when \pagetable pages are placed in NVMM.
We show that (i) \pagetable pages can end up on NVMM even when enough DRAM memory is available and (ii) \pagetable pages that spill over to NVMM due to DRAM memory pressure are not migrated back later when memory 
is available in DRAM.

We study the performance impact of \pagetable placement in a tiered memory system and propose 
an efficient and transparent \pagetable management technique
that (i) applies different placement
policies for data and \pagetable pages, 
(ii) introduces a differentiating policy for \pagetable pages by placing a small but critical part of the \pagetable in DRAM, and 
(iii) dynamically and judiciously manages the rest of the \pagetable by transparently migrating
the \pagetable pages between DRAM and NVMM. 
Our implementation on a real system equipped with Intel's Optane \nvmm
running Linux reduces the \pagetable 
walk cycles by $12\%$ and total cycles by $20\%$ on an average. This improves the runtime by $20\%$ on an average
for a set of synthetic and real-world large memory footprint applications when compared with various default Linux kernel techniques.

\end{abstract}

%% file: introduction.tex
\section{Introduction}
\label{sec:introduction}

The performance of the memory subsystem, both at the software and the hardware layer, is getting
increasingly important in the digital era due to the explosive growth in the amount of data generated,
processed and stored. This along with DRAM scaling challenges~\cite{dram_scale_1,dram_scale_2,dram_scale_3} has led to the exploration of several
new hardware memory technologies with diverse capabilities and capacities such as Intel's Optane PMM
non-volatile main memory (NVMM)~\cite{optane_dc_briefs}.

\begin{figure}
	\centering
	\begin{subfigure}[t]{.55\columnwidth}
	\includegraphics[width=\linewidth]{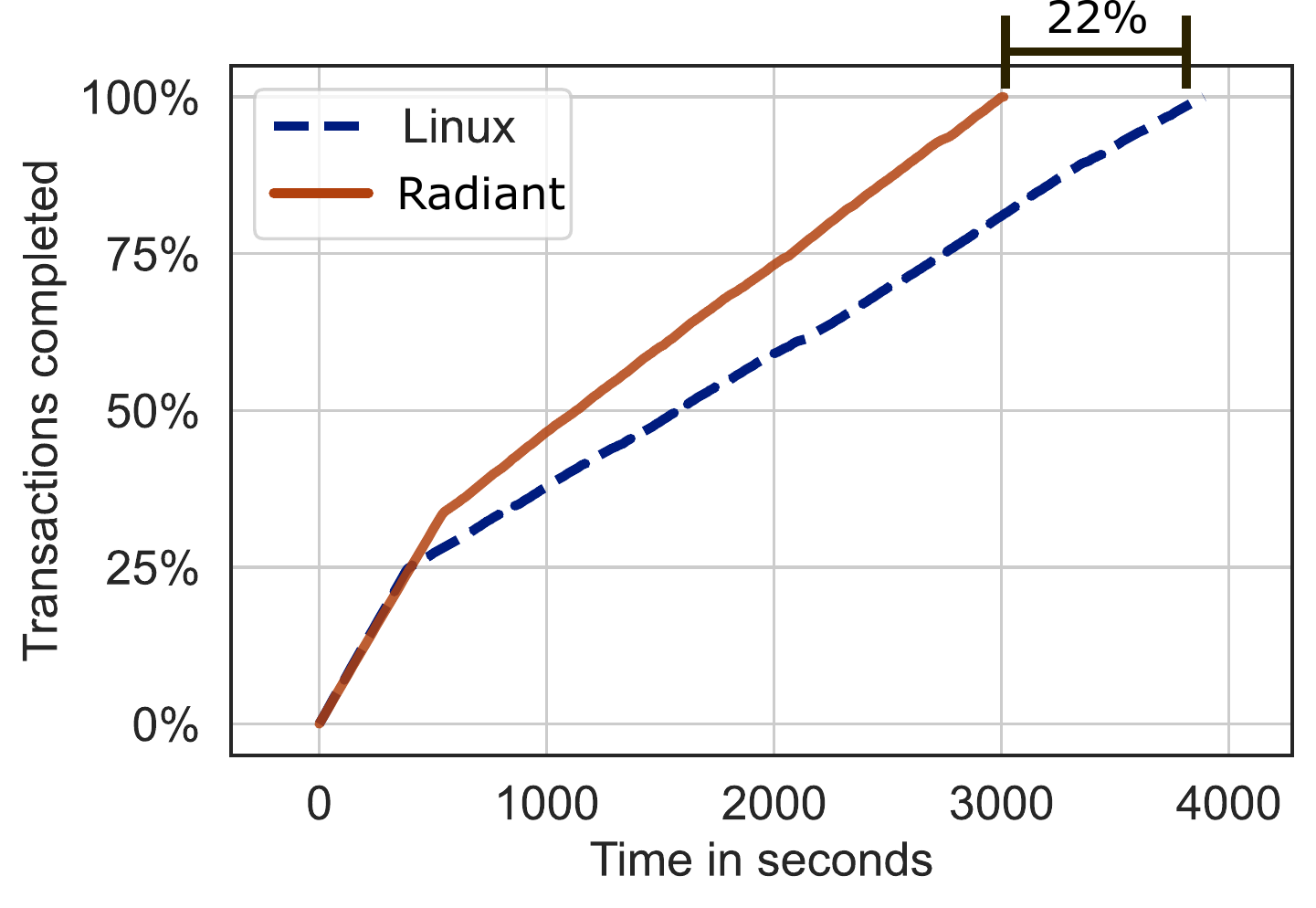}
		\end{subfigure}
	\hfil
		\begin{subfigure}[t]{.4\columnwidth}
	\includegraphics[width=1\linewidth]{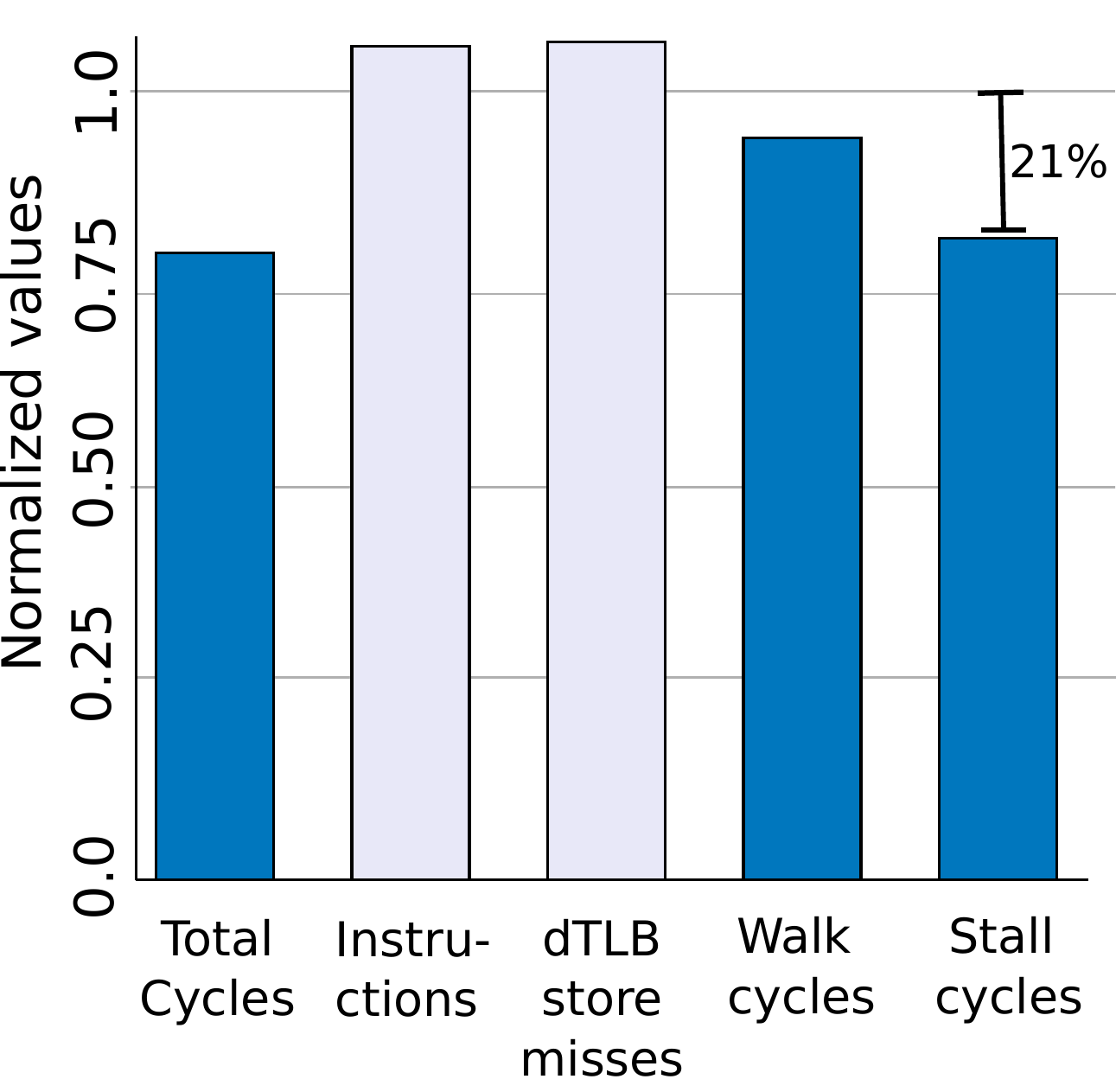}
		\end{subfigure}	
		\caption{Redis populating \(1\,\text{TB}\) of key-value pairs. 
		The inflection at around 500 seconds is when Linux starts allocating 
		both data and \pagetable pages on NVMM. 
		In contrast, \methodname efficiently manages the placement of \pagetable pages 
		between DRAM and NVMM while the data pages follow the Linux kernel's placement policy.}
	\label{fig:default_binded_ops}		
\end{figure}

Modern servers typically use both DRAM and NVMMs to exploit the 
low latency capabilities of DRAM and high capacities of NVMMs~\cite{baidu,optane_graph,autotm}.
Such tiered memory systems bring in additional challenges in terms of managing or tiering 
the placement and the migration of data between DRAM and NVMM. Several prior works~\cite{nimble,bridge_dram_nvmm,traffic_mgmt_numa,memos,heteroos} have studied these challenges for data pages and proposed solutions to identify and migrate hot data pages from NVMM to DRAM. However, they have not studied this in the context of  {\pagetable} pages. We argue that explicit and efficient management
of {\pagetable} pages is crucial for system performance for the following reasons.

\one First, large memory footprint applications, with terabytes of memory, incur frequent
TLB misses~\cite{tlbmiss_bigdata,tlbmiss_bigdata2,tlbmiss_bigdata3} as TLBs cover only a small portion of the total physical memory
(covering few MBs of physical memory with 4K page size and covering up to 3\,GB with 2\,M pages). 
As a consequence, a significant fraction of the memory accesses require a \pagetable walk.

\two Second, the access latency of NVMMs is significantly higher than DRAM. For example, 
on Intel's Optane DC PMM, the read latency is 3$\times$ higher than DRAM, mainly due to 
the Optane's longer media latency~\cite{fastpaper}. Consequently, a hardware \pagetable walk incurs higher
walk latency when a \pagetable page is placed in NVMM. As a \pagetable walk requires
up to 4 memory accesses upon a TLB miss (for a 4-level \pagetable), the \pagetable 
walk latency can be significantly higher in such cases which negatively impacts the application's performance (as shown in Figure~\ref{fig:default_binded_ops}).
\methodname efficiently places the \pagetable pages
between DRAM and NVMM to reduce cycles spent in page table walks which in turn improves the start-up time of Redis by 22\% (Figure~\ref{fig:default_binded_ops}).

\three Third, a typical \pagetable occupies a small fraction of DRAM. For example, the \pagetable size of an application with 2\,TB memory footprint is around 4\,GB which is around 1\% of DRAM on our evaluation system. Despite its relatively small size, page table pages can end up on NVMM even when there is enough free memory in DRAM. For instance, 
existing operating systems do not differentiate between \pagetable 
and data page allocations; they apply the same allocation policy for both of them~\cite{mitosis,numa_challenges,traffic_mgmt_numa}. Hence,
when memory interleave policy is selected for data pages, \pagetable pages are also allocated in a round robin order on all nodes, including NVMM nodes, even when DRAM has free memory.

\four Lastly, operating systems do not support migration of \pagetable pages~\cite{mitosis}. 
Once the \pagetable pages are allocated, they remain fixed for their lifetime; they are reclaimed
only when either the corresponding data pages are freed or the process is terminated.
In contrast, data pages enjoy the flexibility of migration between DRAM and NVMM based on the application's memory access pattern.

A simple and straight forward approach to avoid \pagetable pages spilling to NVMM is
to bind the \pagetable to DRAM. However, this approach results in pathological behavior
where applications are killed by the out-of-memory (OOM) handler even when significant amount of free memory is available in the system (details in \S\ref{motivation:binding}). In addition, as all the \pagetable pages are not frequently accessed, placing the complete \pagetable on high-performance DRAM memory is not merited.
Hence, we argue for judiciously managing the placement of \pagetable pages across
DRAM and NVMM.

In this paper, we propose \methodname, an efficient and transparent \pagetable
management technique for tiered memory systems. 
\methodname differentiates between a data and a \pagetable page allocation by applying 
different placement policies to them. It also considers the
underlying memory heterogeneity while deciding on the placement of the \pagetable pages.

Additionally, \methodname employs the following techniques for efficient
\pagetable management:
\begin{itemize}
	\item {\bf Placement:} introduces a differentiating placement policy within the \pagetable by placing a small but critical part of the \pagetable in DRAM. 
	This differentiating placement strategy is based on the observation that the top three levels of 
	a \pagetable tree forms a small portion of the \pagetable but are frequently accessed
	during a \pagetable walk (3 out of 4 accesses during a page walk are from the higher levels of a \pagetable). 

	\item {\bf Migration:} efficiently identifies and transparently migrates the last level \pagetable pages between memory tiers by employing a novel data-\-page-\-migration triggered \pagetable migration technique.
\end{itemize}

We implement \methodname in the Linux kernel and evaluate the performance benefits
on a real system equipped with Intel's Optane PMM persistent memory. 
\radiant reduces the \pagetable walk cycles by $12\%$ and total cycles by $20\%$ on an average. 
This improves the runtime by $20\%$ on an average
for a set of synthetic and real-world large memory footprint applications when compared with the various default Linux kernel techniques.

The main contributions of the paper are as follows:
\begin{itemize}

	\item Based on extensive characterization and experimentation on a diverse set of workloads, we argue that different placement and migration policies are required for data and \pagetable pages in tiered memory systems.
	\item To the best of our knowledge, this is the first work that focuses on
		efficient placement and migration of {\pagetable}s on tiered memory systems.
	\item A differentiating placement policy within the \pagetable where a small but
	critical part of \pagetable pages are allocated on DRAM while the rest of the \pagetable pages are dynamically managed by migrating between memory tiers.
\end{itemize}

The rest of the paper is organized as follows: we provide the necessary background in Section~\ref{sec:background} followed by the motivation for the paper in Section~\ref{sec:motivation}. We present our design in Section~\ref{sec:design} and implementation details in
Section~\ref{sec:implementation}. We evaluate the performance of \radiant in Section~\ref{sec:evaluation}. We briefly discuss related works in Section~\ref{sec:relatedworks} and finally, conclude in Section~\ref{sec:conclusion}.

%% file: background.tex
\label{sec:numa}

\section{Background}
\label{sec:background}

In this section, we cover the necessary background required for the rest of the paper.

\subsection{Optane Persistent Memory}
Intel's Optane Persistent Memory Module is a high-capacity non-volatile memory that 
is DDR4 socket compatible and fits into standard DIMM slots~\cite{optane_dc_briefs}.
Optane can be used either as a high-capacity volatile main memory (Memory Mode 
and Flat Mode)
or as a persistent memory (App Direct Mode)~\cite{fastpaper,basic_optane_measurement,hpc_optane_measurement}. 
High capacity volatile memory 
is useful for large memory footprint applications as they can
exploit the advantage of the additional memory capacity without
requiring application modifications. For example, Optane
can seamlessly enable large-scale in-memory graph analytics for graphs with billions of edges ~\cite{optane_graph}.

In this work, we use Optane as a high-capacity volatile memory in 
Flat Mode (also referred to as DRAM-NVMM hybrid mode~\cite{hpc_optane_measurement}).
The difference between Memory Mode and Flat Mode is that 
in Memory Mode, Optane acts as a byte-addressable volatile
main memory while DRAM acts as a cache.
In Flat Mode, both DRAM and Optane memory can be accessed as a unified, but heterogeneous, byte-addressable memory.
The advantage with Flat Mode is that the applications can control and optimize the
placement of data between low latency DRAM and high latency Optane~\cite{baidu,autotm}.

\begin{figure}
	\centering
	\includegraphics[scale=0.73]{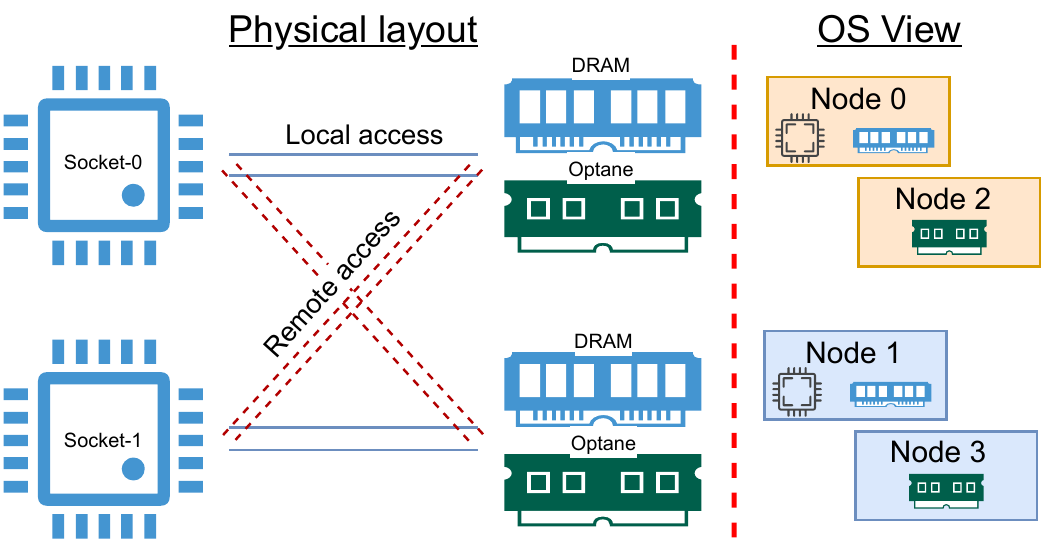}
	\caption{A 2-socket system equipped with Intel's Optane memory. The two sockets are logically divided into four NUMA nodes in Linux.  Node 0 and Node 1 are backed by DRAM while Node 2 and Node 3 are backed by Optane.}
	\label{fig:socket_placement}
\end{figure}

We configure the system in Flat Mode using \texttt{ndctl} tool~\cite{ndctl}
and \texttt{daxctl} utility~\cite{daxctl}. Step by step guide
to configure Optane as a hot-plugged main memory is available
in Persistent Memory Development Kit (PMDK)~\cite{flat_mode}.
Once configured in Flat Mode, Optane memory is reflected as ``no-CPU'' NUMA nodes in the
system as shown in Figure~\ref{fig:socket_placement} (node 2 and node 3).
Support for Flat Mode is already
part of the Linux kernel~\cite{kernel_support_nvmm} and hence
all the NUMA features (e.g., placement and balancing) in Linux are
readily available for Optane-backed NUMA nodes as well.

\subsection{Page tables}

\begin{figure}
	\centering
	\includegraphics[width=.9\columnwidth]{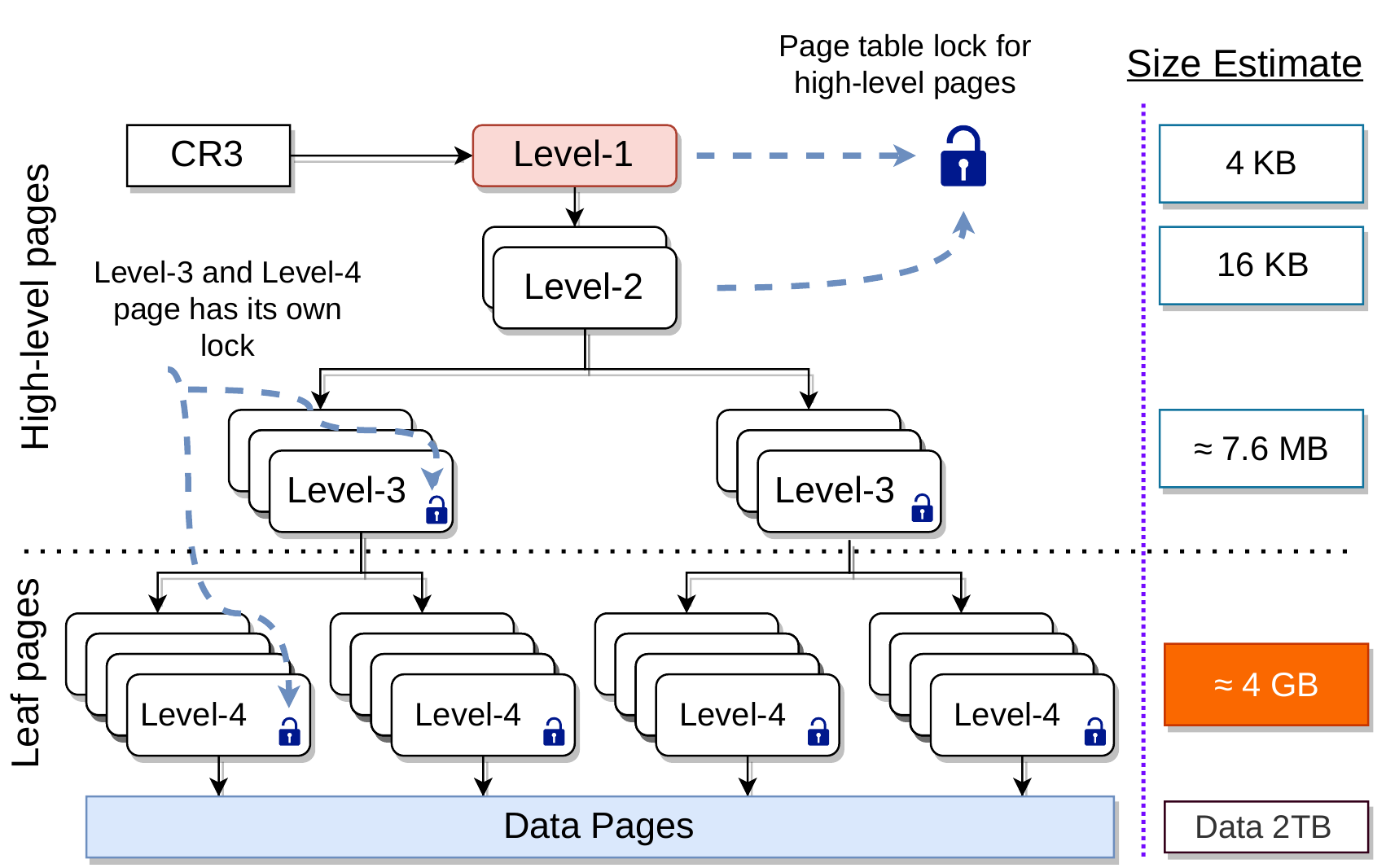}
	\caption{Figure depicting the structure of a 4-level \pagetable.}
	\label{fig:page_table}
\end{figure}

A \pagetable maintains virtual address (VA) to physical address (PA) translations and is organized as a multi-leveled tree (\texttt{x86\_64} supports both 4-level and 5-level {\pagetable}s; we use 4-level \pagetable for the discussions in the rest of the paper\footnote{A 4-level \pagetable can map up to 256\,TB of memory.}) where a 	page global directory (PGD or L1) is the root of the tree. Each active entry in PGD points to a physical page containing an array of page upper directory (PUD or L2) entries. Similarly, each active entry in PUD points to a physical page containing an array of page middle directory (PMD or L3 ) entries. PMDs in turn point to a physical page (PTE or L4) containing an array of \pagetable entries. A PTE entry contains the physical page address of the data page corresponding to the virtual address as shown in Figure~\ref{fig:page_table}.

Upon a CPU TLB (Translation Lookaside Buffer) miss, the hardware -- being aware of the \pagetable tree layout -- performs a \pagetable walk to insert an entry in the TLB. Therefore, a TLB miss on a 4-level \pagetable requires 4 memory accesses to walk the \pagetable. As TLBs cover only a small portion of the total physical memory, most of the memory accesses by large memory footprint workloads cause a TLB miss requiring a \pagetable walk. 

In modern operating systems, {\pagetable}s are dynamically allocated: the root of the \pagetable tree for a process is allocated when the process is created. The physical pages to store the intermediate and leaf-level pages of the \pagetable are allocated whenever the process page-faults on a valid virtual address for the first time.

\subsection{Userspace data page allocation and migration}

Modern operating systems such as Linux provide a stable and transparent technique for data page allocation on a multi-socket system. Additionally, they also provide mature interfaces or APIs for applications to explicitly control data page allocation. By default, Linux employs a first-touch policy~\cite{numa_challenges,mitosis}, which allocates data pages on a local NUMA node and falls back to remote nodes when there is
not enough memory on the local node. Apart from this, an interleaved allocation policy~\cite{numa_challenges} is also available where the data pages
are allocated on all NUMA nodes in a round robin order. This improves memory bandwidth utilization by distributing the data pages across nodes and thus, avoids skewed allocation to a set of nodes~\cite{numa_challenges}.

In a NUMA system, 
accessing data from a remote node causes significant memory overheads incurring 2--4$\times$ higher latency than accessing the data from a local node~\cite{mitosis}. Many solutions have been proposed over the last few decades to mitigate such performance issues~\cite{traffic_mgmt_numa,numa_placement1,numa_placement2}. One of the solutions that is widely accepted and used is to migrate the data pages from the remote NUMA node to a local NUMA node where the application is running. 

Operating systems such as Linux provides well defined userspace APIs to trigger data page migrations between NUMA nodes~\cite{move_pages}. In addition, operating systems are capable of transparently migrating frequently accessed data pages between NUMA nodes (e.g., AutoNUMA in Linux~\cite{autonuma}). 
However, it is important
to note that the page migration support is only available for userspace data pages; such support is not available for kernel pages and consequently, \pagetable pages.

%% file: motivation.tex
\section{Motivation}
\label{sec:motivation}

In this section, we present a \pagetable analysis for large memory footprint applications including the placement and distribution
of \pagetable pages, migration of \pagetable pages and performance impact of \pagetable placement. 
We draw important observations based on our analysis and
design \methodname based on these observations and insights.

We run our experiments on a 2-socket Intel-Xeon Gold 6252N system with 192\,GB DRAM and 800\,GB Optane per socket
running Linux kernel version 5.6 (system and configuration details in Table~\ref{tab:system_config}).
 
\subsection{TLB misses}
\label{motivation:mpki}
\begin{figure}
	\centering
	\includegraphics[scale=0.45]{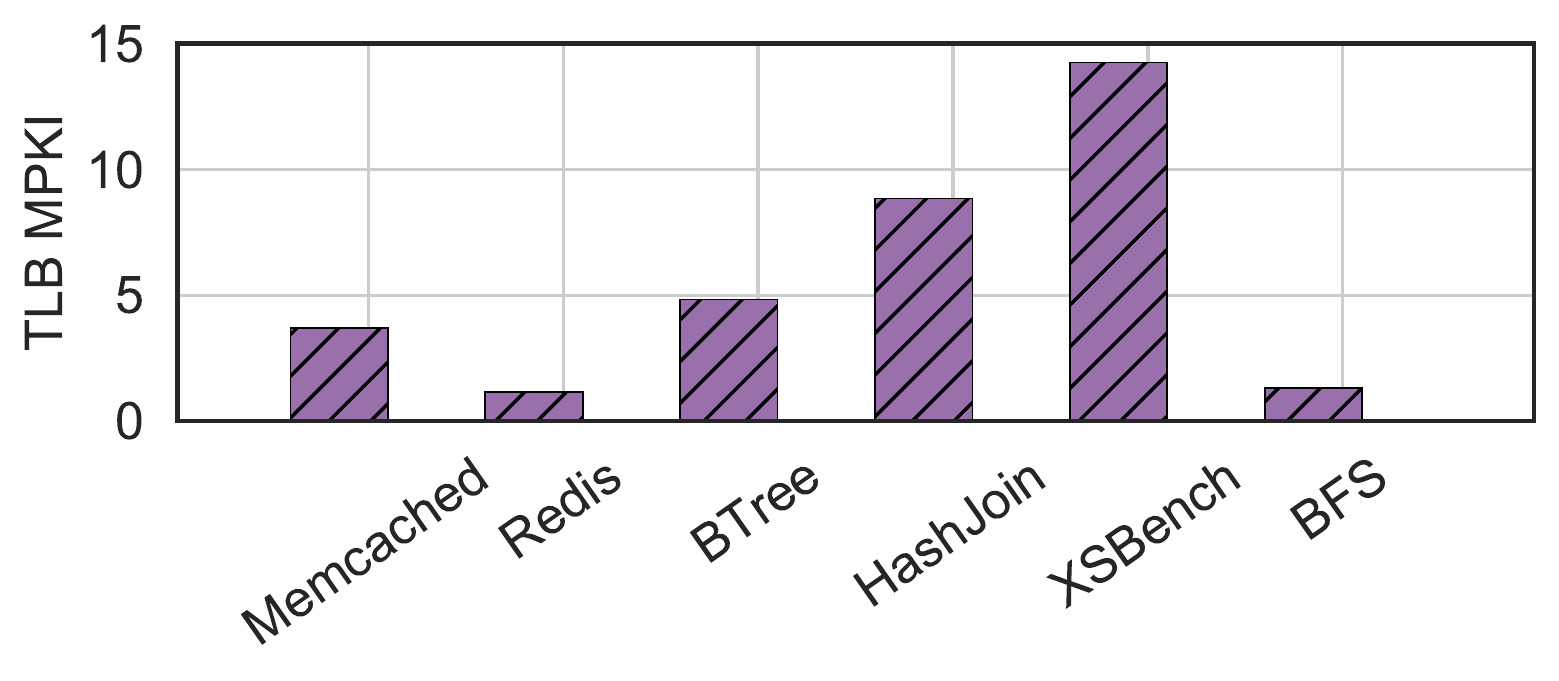}
	\caption{TLB MPKI for applications with large memory footprint. Benchmark details in Table~\ref{tab:bench}.}
	\label{fig:motivation_mpki.pdf}
\end{figure}
Large memory footprint applications running terabytes of memory usage incur frequent
TLB misses as TLBs cover only a small portion of the total physical memory. 
Figure~\ref{fig:motivation_mpki.pdf} shows the TLB Misses-Per-Kilo-Instructions (MPKI)
for applications with large memory footprint (600\,GB to 1\,TB). 
A higher MPKI implies that a significant fraction of the memory accesses 
incurs TLB misses, thus requiring \pagetable walks. 

Due to high TLB misses, these applications spend up to 68\% of the total execution cycles in \pagetable walks. 
It is important to note that MMU employs caching techniques to cache
the \pagetable entries. Additionally, \pagetable entries are also 
cached in system memory caches as MMU units access the \pagetable 
through the memory hierarchy. Despite MMU caching and other TLB 
optimization techniques, large memory footprint applications still
spend significant fraction of the total execution cycles
in \pagetable walks.
Hence, optimizing \pagetable walks are important for such applications.

\subsection{Page table placement}
\label{motivation:placement}
Operating systems dynamically allocate pages for all the four levels of \pagetable on-demand, i.e., when the corresponding virtual address page faults for the first time. However, the NUMA node on which a \pagetable page
is allocated depends on multiple factors including the socket on which the allocating thread is running and the memory allocation policy of 
the application~\cite{numa_challenges,traffic_mgmt_numa}. It is important to
note that operating systems employ the same allocation and placement policy for both data and \pagetable pages.

\begin{figure}
	\centering
	\includegraphics[width=.9\columnwidth]{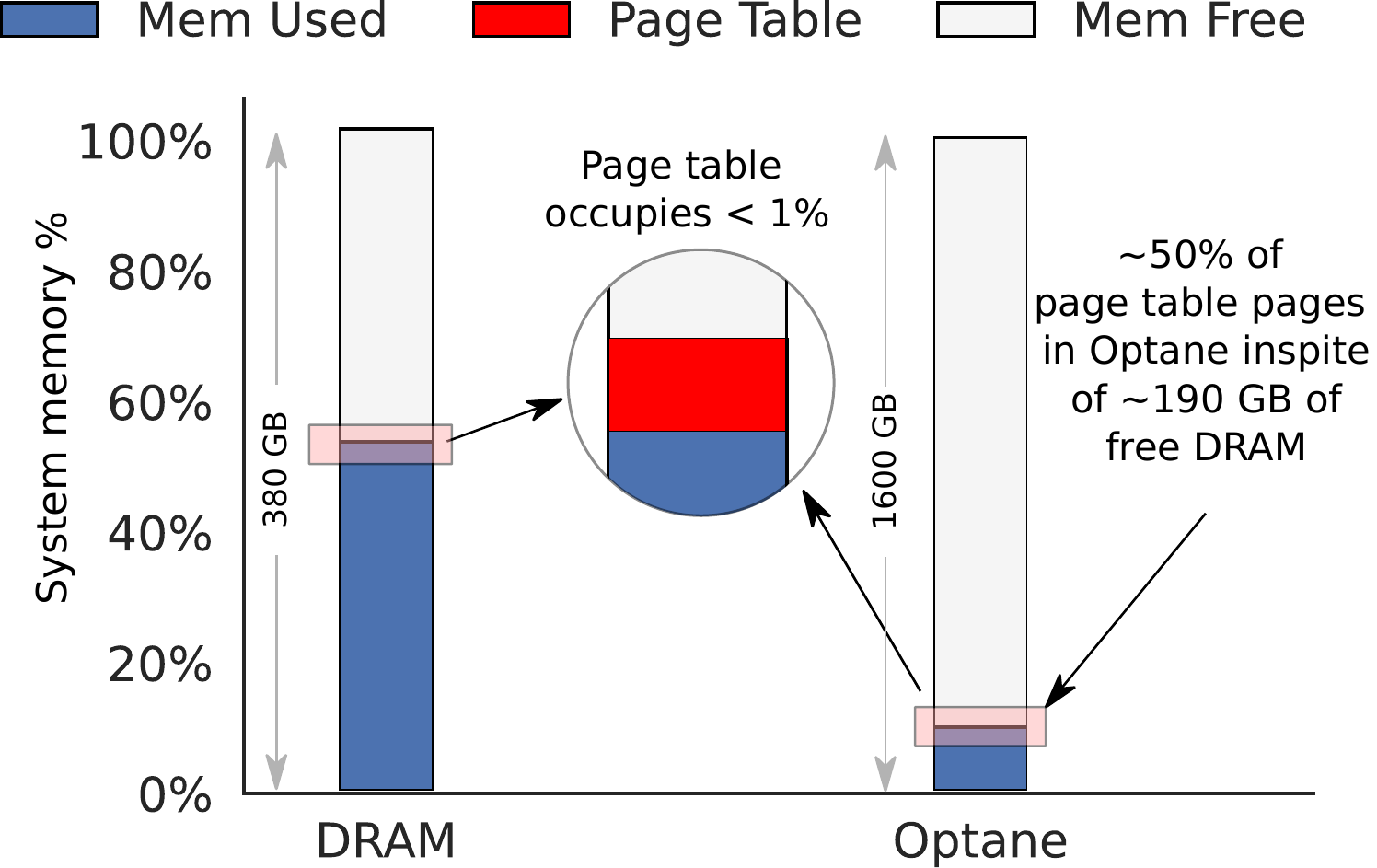}
\\
\scriptsize
\begin{tabular}{ccccc}
            & & & & \\
			\toprule
			 & \multicolumn{2}{c}{\textbf{ DRAM}} & \multicolumn{2}{c}{\textbf{Optane}} \\
			Memcached stats & Node 0 & Node 1 & Node 2 & Node 3 \\
			\midrule
			Data (in GBs) & 79.00 & 90.00 & 84.00 & 84.00 \\
			Page table (in GBs) & 00.17 & 00.17 & 00.16 & 00.16 \\
			\bottomrule
		\end{tabular}

	\caption{Page table distribution for \memcached when around 338\,GB of data has been populated with interleaved allocation policy. Around $50\%$ of page table pages end up in \nvmm even when 190\,GB of DRAM is free.}
	\label{fig:styl_1.pdf}
\end{figure}

When the memory interleave policy 
(round-robin allocation of pages across all NUMA nodes) is applied for an application,
the Linux kernel applies the same data page allocation policy for the \pagetable pages also and allocates them across all NUMA nodes. Figure~\ref{fig:styl_1.pdf} shows the
placement of \pagetable pages and data pages when around 338\,GB of data has been populated in Memcached. It can be observed that around 50\% (0.32\,GB) of
page table pages are allocated in Optane despite having around 190\,GB free memory in DRAM.

Furthermore, when first-touch allocation policy is applied, allocation of \pagetable pages spills over to Optane when DRAM is almost full.
However, later when a part of DRAM memory is freed, data pages are migrate from Optane to DRAM. However, \pagetable pages remain in Optane as they cannot be migrated.

As a result, in one scenario, \pagetable pages can be allocated in NVMM even when enough free memory is available in DRAM and in another scenario \pagetable pages allocated on NVMM remains on NVMM even when enough memory is freed on DRAM (\textbf{Observation 1}). 

\subsection{Page walk latency}
\label{motivation:walk_latency}
\begin{figure}
	\centering
		\includegraphics[scale=.4]{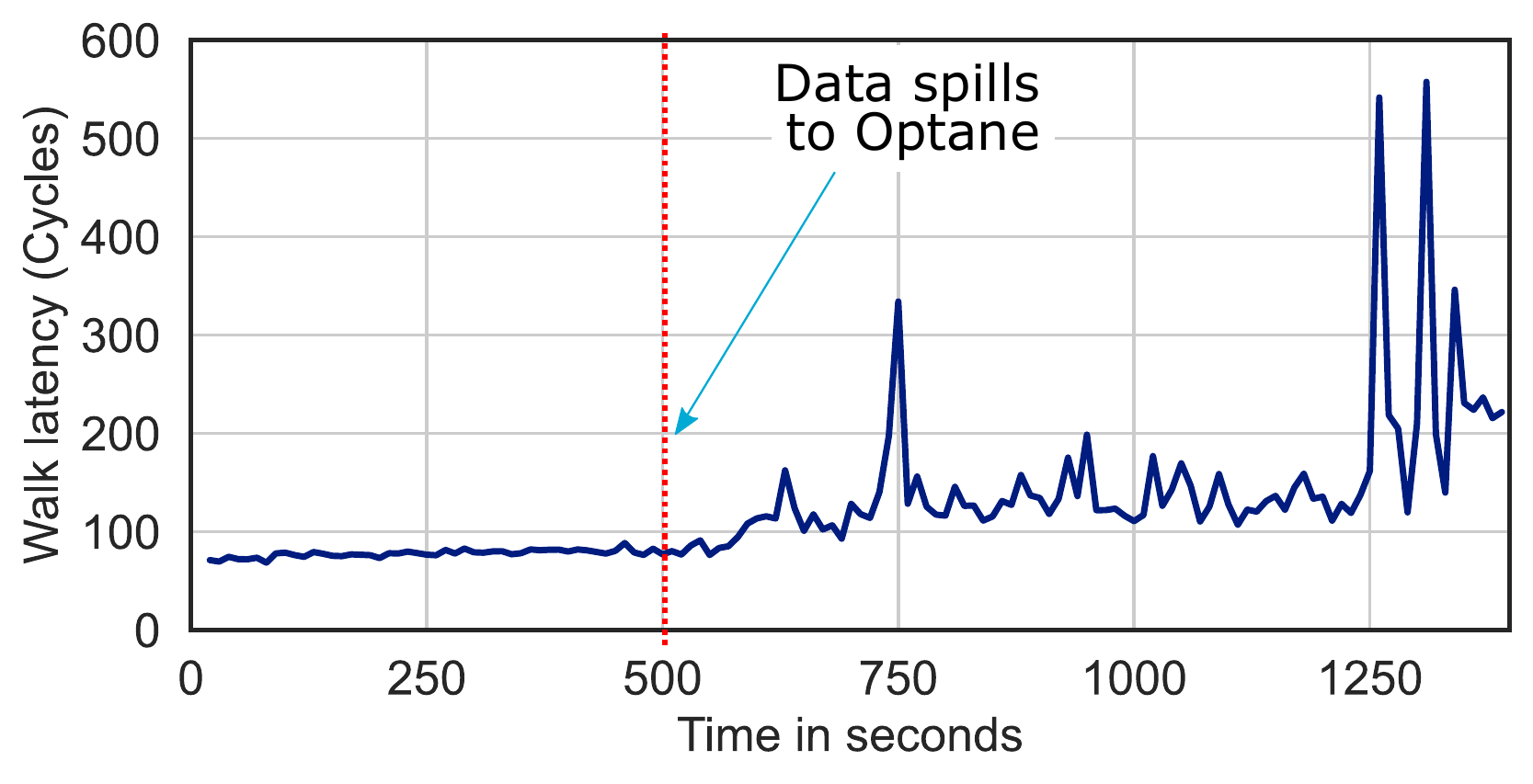}
	\caption{Page walk latency when populating Redis with 1\,TB of key-value pairs (we plot the first 1400 seconds, but the trend continues).
	Page walk latency increases when the \pagetable page allocation spills to Optane.}
	\label{fig:motivation_walk_latency}
\end{figure}

The access latency of NVMMs are significantly higher than DRAM mainly due to the 
longer media latency. Hence, a hardware \pagetable walk incurs higher
walk latency when a \pagetable entry is placed in NVMM. 
Additionally, a \pagetable walk requires up to 4 memory accesses to NVMM
when all the four levels of page table pages are allocated in NVMM. This
further increases the page walk latency. 
It has also been observed that concurrent access to NVMMs, especially Optane, 
from multiple CPUs in a multi-core system can degrade performance due to
limited internal buffers~\cite{fastpaper}.   
 
We measure the page walk latency when populating Redis with 1\,TB of 
key-value pairs.
We use the default placement policy (first-touch policy) which allocates
data and \pagetable pages from DRAM until the free memory in DRAM
reaches a critical mark before falling back to NVMM.
Page walk latency increases significantly (Figure~\ref{fig:motivation_walk_latency})
when the \pagetable page allocation spills to NVMM (\textbf{Observation 2}).

\subsection{Migration support}
\label{motivation:migration}
Techniques employed by operating systems and userspace applications to identify and migrate frequently accessed pages from NVMM to DRAM to improve application performance are restricted to data pages.
To the best of our knowledge, most modern operating systems do not support migrating a \pagetable page because
\pagetable pages are part of the kernel pages, which are unmovable. 
Once the \pagetable pages are allocated, they remain fixed for their lifetime; they are reclaimed only when either the corresponding data pages are freed or the process is terminated. As a result, \pagetable pages allocated on NVMM remain in NVMM.

Migrating a \pagetable page is a non-trivial operation as it
requires fixing
the \pagetable tree structure (a kernel data structure) to ensure that the virtual to physical 
address mappings are intact.
In addition, \pagetable page migration on a multi-core system requires 
careful handling of race conditions.
For example, the \pagetable page under migration can either be accessed by hardware during a page walk or can be accessed/modified by other CPUs to serve a page fault.

\subsection{Page table binding}
\label{motivation:binding}
\begin{figure}[ht]
	\centering
	\includegraphics[scale=0.4]{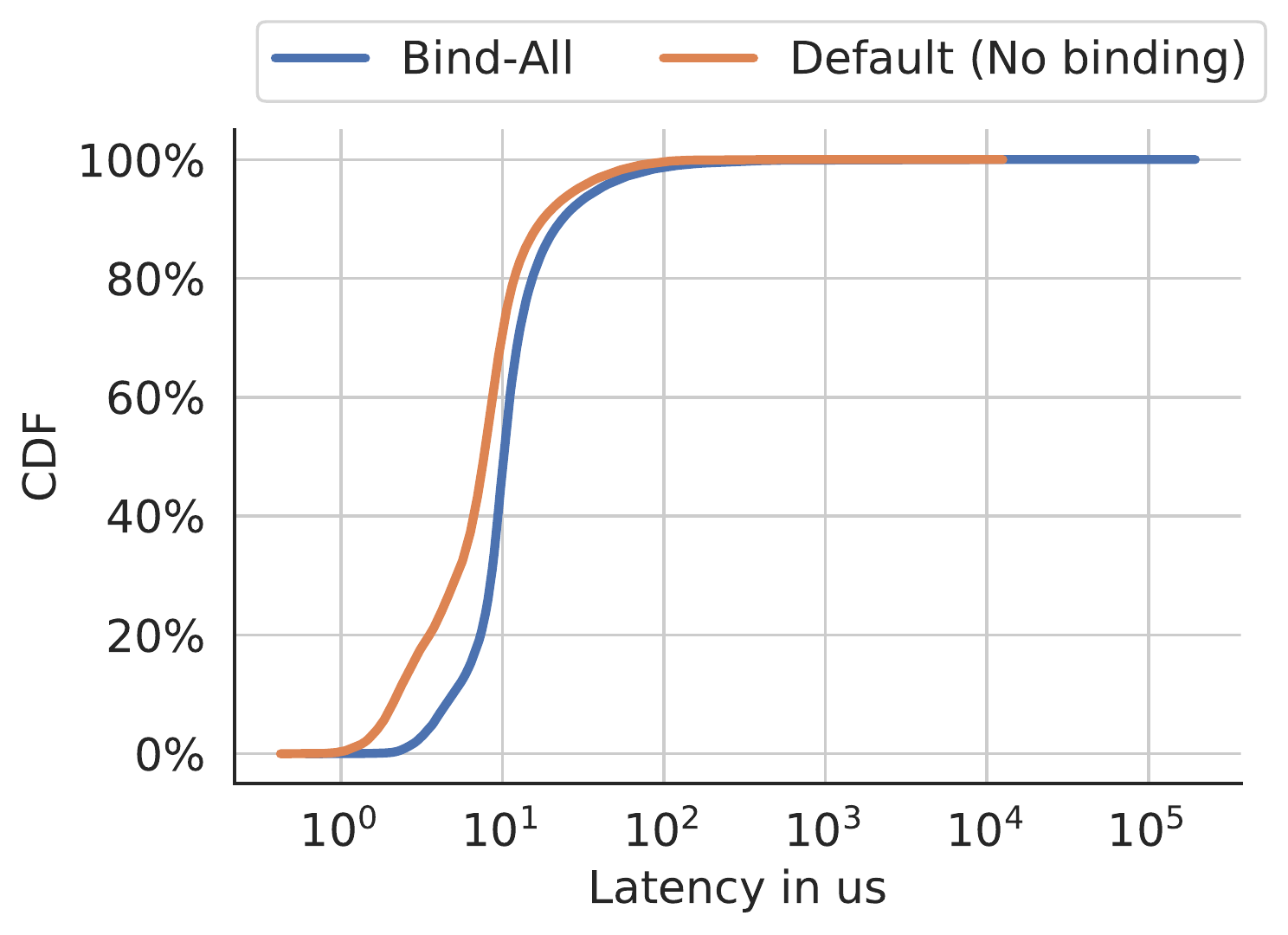}
	\caption{L4 \pagetable page allocation latency in case of the default Linux kernel and when the entire \pagetable is binded to DRAM. }
	\label{fig:pte_alloc_latency}
\end{figure}
A simple and straight forward approach to avoid \pagetable pages spilling to NVMM is to bind the \pagetable to DRAM. Even though this looks like a viable options, it results in pathological behaviour as we demonstrate by evaluating the Linux kernel patches~\cite{linux_pgtable} that propose to bind 
the \pagetable to DRAM.\footnote{These patches are not included in the Linux kernel;  Linux kernel v5.6 still allows allocation of \pagetable pages on Optane NUMA nodes.}

On a freshly booted system with 384\,GB DRAM and 1.6TB Optane memory, we start populating Memcached
 with the default first-touch allocation policy.
Initially, all new page allocations (both data and page table pages), as per first-touch policy, are directed to DRAM resulting in DRAM nodes filling up before Opatne nodes (on our system DRAM is 19\% of the total system memory). 
Once DRAM is almost full, new data page allocations are directed to Optane nodes, while the \pagetable pages are still directed to DRAM due to DRAM binding. 
Forcing the allocation on almost-full DRAM nodes results in higher allocation latencies (Figure~\ref{fig:pte_alloc_latency}) because the buddy allocator falls back to slowpath
functions. 

In addition, continued \pagetable allocation requests on almost-full DRAM nodes invokes page reclamation daemon. However, after a while, the Linux kernel fails to reclaim enough DRAM pages to serve \pagetable page allocation and as a result triggers the out-of-memory (OOM) handler. OOM handler kills the Memcached server even when 700\,GB of free memory is available in Optane NUMA nodes.

Out-of-memory issues can be mitigated to some extent by employing aggressive page reclamation heuristics, however, we address these challenges by efficiently handling the allocation and placement of \pagetable pages across memory tiers. Our approach is to allow \pagetable pages to spill over to NVMMs when allocation on DRAM is not possible and then dynamically and transparently migrate the \pagetable pages between the memory tiers.

\subsection{Summary}
To summarize, we argue that with the growing relevance of large tiered memory systems,
it is important to explore efficient \pagetable
allocation and placement technique across memory tiers, which has received least attention till now.

%% file: design.tex
\section{\methodname design}
\label{sec:design}

We propose an efficient and transparent \pagetable management
technique to reduce page walk overheads on 
tiered memory systems. In this section, we present the design
of \methodname.

\subsection{Design considerations}

\noindent {\bf Differentiate between data and \pagetable pages:}
Large memory footprint applications with terabytes of memory incur frequent TLB misses. 
The performance of such applications is sensitive to the placement of \pagetable pages in a
tiered memory system. Hence, it is necessary to consider different allocation and placement
policies for data and \pagetable pages.

\smallskip

\noindent {\bf Differentiate between NVMM and DRAM memory:}
Carefully consider the underlying memory heterogeneity (e.g., capacity, latency)
while deciding on the placement of \pagetable pages.

We propose the following two techniques that incorporate the above design considerations along with
the observations made during \pagetable analysis in \S\ref{sec:motivation}.

\subsection{Binding critical \pagetable pages to DRAM}

The read latency on NVMM is 3$\times$ higher than DRAM mainly due to the longer media latency.
As a \pagetable walk requires 4 memory accesses, the \pagetable
walk latency is significantly higher when all the four levels of the \pagetable pages are allocated on
NVMM. Even though a typical \pagetable for a 
large memory footprint application can occupy a small fraction of DRAM,
binding the entire \pagetable to DRAM can result
in pathological behaviour as demonstrated in \S\ref{motivation:binding}.

We observe that a majority of the \pagetable memory 
is consumed by leaf level or L4 \pagetable pages; 
L1, L2 and L3 \pagetable pages together consume insignificant amount of memory.
For example, an application with around 2\,TB memory footprint requires around 4\,GB
memory for L4 pages and collectively requires around 7.62\,MB for L1, L2 and L3 
\pagetable pages (size estimation in Figure~\ref{fig:page_table}).
We exploit this insight to significantly reduce the amount of time spent on \pagetable walks. 

Our placement strategy is to dynamically allocate and bind L1, L2, and L3
\pagetable pages in DRAM. 
With such a placement technique, during a 4-level page walk, 3 out of 4 memory accesses are guaranteed from low latency DRAM thus drastically reducing the page walk cycles.
It is important to note that we achieve this by strategically placing 
less than 0.18\% of \pagetable pages in the critical DRAM memory.

Such a policy not only improves the application execution time 
but also improves startup or initialization time for large memory footprint 
applications. For example, when populating initial key-values in an in-memory database,
initializing a large graph or restoring a VM snapshot, a large portion of L1, L2, and L3
\pagetable pages are initialized and accessed (e.g., zeroing a newly allocated
\pagetable page). Hence, placing them in DRAM reduces the startup time of applications.

Our strategy, as opposed to placing the entire \pagetable in DRAM~\cite{linux_pgtable}
has several advantages. \one First, we drastically minimize the amount of \pagetable pages that requires binding to DRAM. For example, we bind only 7.62\,MB for a 2\,TB workload which is less than 0.0019\% of DRAM on our evaluation system. In contrast binding the entire \pagetable requires 4\,GB of DRAM. \two Second, by using less than 0.0019\% of DRAM for binding we guarantee 75\% of \pagetable walks from DRAM. \three Finally, even under extreme memory pressure operating systems can allocate L1, L2 and L3 \pagetable pages in DRAM by reclaiming a small amount of DRAM memory. While binding the entire page table requires reclaiming few GBs of DRAM memory which can trigger out-of-memory handler.

\subsection{Page table migration}

We allow allocation of L4 \pagetable pages, which constitutes the majority of the \pagetable, on both DRAM and NVMM. Further, we use 
data-page-migration triggered \pagetable migration technique
to efficiently identify and migrate L4 pages between DRAM and NVMM.

The rational behind such an approach is that a data page migration
provides crucial hint on the placement of the corresponding L4 \pagetable
page. For example, migration of a hot data page from NVMM to DRAM hints that
the corresponding L4 \pagetable page, if present on NVMM, should also be migrated.
Because, for a large memory footprint application with terabytes
of memory even a hot data page incurs frequent TLB misses
(as the amount of hot data far more exceeds the TLB reach) 
resulting in frequent accesses to L4 page by the hardware page walker.
Therefore, when a data page is migrated between memory
tiers we trigger the migration of the corresponding L4 \pagetable page.

Operating systems such as Linux provides well defined userspace API~\cite{move_pages}
to trigger data page migrations to enable novel userspace 
techniques to efficiently identify and migrate data pages between memory
tier. For example, identifying and migrating hot and cold data pages between memory tiers or
speculatively pre-migrating a set of data pages between DRAM and NVMM based on the application's memory access patterns. In addition, operating systems are capable of transparently migrating
frequently accessed data pages between NUMA nodes (e.g., AutoNUMA in Linux). 
We exploit such existing data migration techniques to
trigger an L4 \pagetable page migration between DRAM and NVMM.

We migrate an L4 page from NVMM to DRAM upon the migration of the corresponding data page, 
however, we migrate an L4 page from DRAM to NVMM only when
the last data page it is pointing to is migrated to NVMM.
This is to ensure that an L4 page is in DRAM if 
any data page it is pointing to is in DRAM.

\newcommand{\olddp}{old data page\xspace}
\newcommand{\newdp}{new data page\xspace}
\newcommand{\pte}{PTE\xspace}

\newcommand{\algodp}{$data\_page$\xspace}
\newcommand{\algoolddp}{$data\_page_{old}$\xspace}
\newcommand{\algonewdp}{$data\_page_{new}$\xspace}
\newcommand{\algodestnode}{$dest\_node$\xspace}

\subsection{Page table migration details}
\label{sec:migrating_pte_page}

\begin{figure}[!ht]
	\hspace*{-.3cm}         
	\centering
	\includegraphics[width=\linewidth]{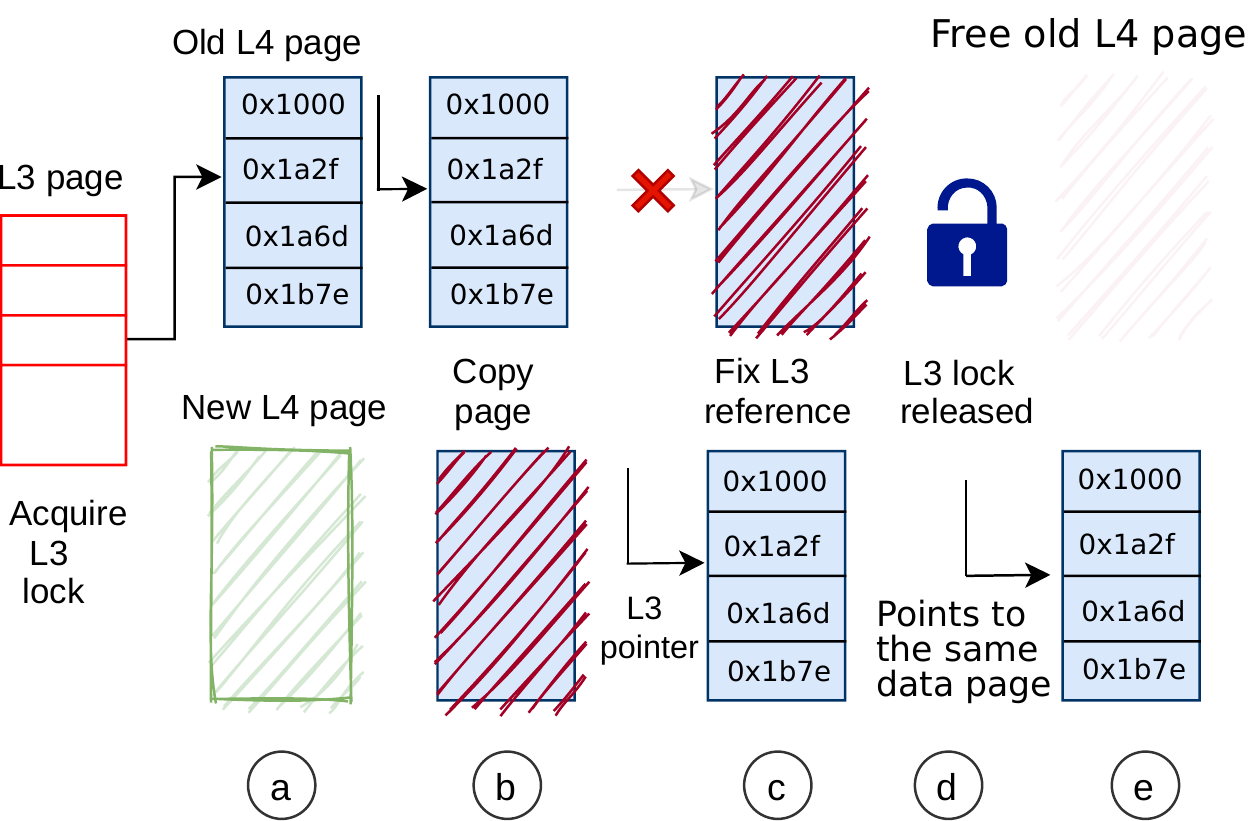}
	\caption{Steps for a \pagetable page migration. A shaded page indicates an un-referenced page.}
	\label{fig:default_page_migration}
\end{figure}

\begin{algorithm}
	\footnotesize
	\caption{Algorithm to migrate an L4 page}
	\label{algo:pte_migration}
	\begin{algorithmic}[1]
		\Procedure{Migrate\_Data}{\algodp, \algodestnode}
		\State \algonewdp $\gets$ \texttt{alloc\_page(\algodestnode)}
		\State rc$\gets$\texttt{migrate\_data\_page}(\algodp, \algonewdp, \algodestnode)
		\If{rc==SUCCESS} \label{algoline:dpmigrate_success}
		\State migrate\_L4(\algonewdp, \algodestnode) \Comment{Migrate L4 page}
		\EndIf
		\EndProcedure
		
		\\\hrulefill
		\Procedure{migrate\_l4}{\algonewdp, \algodestnode}
		
	\State \textcolor{blue}{ 	\textit{/* Walk the page table */}}
		\State $(L4,L3) \gets$ \texttt{get\_pt\_entries(\algonewdp)} 
		\label{algoling:getmappings}
		
		\State $L4\_node \gets$ \texttt{page\_node($L4$)} \Comment{Get L4's node}
		\If {$L4\_node == dest\_node$}  
		\State return 
		\Comment{Already in destination}
		\label{algoline:indest}
		
		\ElsIf {$L4\_node$ in DRAM and \algodestnode in DRAM} 
		\State return 
		\Comment{Already in DRAM (similarly for NVMM)}
		\label{algoline:indram}
		\ElsIf {$dest\_node$ in \nvmm} 
		\If {\textbf{any} data page pointed by $L4$ in DRAM}
		\State return 
		\Comment{L4 pointing to a page in DRAM}
		\label{algoline:hotdata}
		\EndIf
		\EndIf
		
		\If {\texttt{lock}($L4$ and $L3$)}
		\Comment{Lock \textit{L4} and \textit{L3} pages}
		\label{algoline:lockpages}
		
		\State \textcolor{blue}{\textit{/*Allocate a \textit{L4} page on the destination NUMA node*/} }
		\State $ L4_{new}\gets$ \texttt{alloc\_page}(dest\_node) 
		\label{algoline:allocpte}
		
		\State \texttt{tlb\_flush()}
		\Comment{Invalidate $L4_{old}$ mappings}
		\label{algoline:flush_1}

		\State \texttt{memcpy}($ L4_{new} $, $ L4$, 4096) 
		\Comment{Copy the \textit{L4} page}
		\label{algoline:memcpy}
		
		\State \texttt{update\_L3}($ L4_{new} $)  \label{algoline:sync_pt} 
		\Comment{Sync. point}
		\label{algoline:sync-pt}
		
		\State \texttt{unlock}($L3$ and $L4$)
		\Comment{Unlock the \textit{L3} and \textit{L4} pages}
		\label{algoline:unlockpages}
		\EndIf
		\EndProcedure
		
	\end{algorithmic}
\end{algorithm}

Algorithm~\ref{algo:pte_migration} and 
Figure~\ref{fig:default_page_migration} show the steps
involved in migrating an L4 \pagetable page.
Whenever a data page migration is
initiated either by an userspace program or by the kernel (e.g., AutoNUMA),
we trigger the migration of the corresponding \pagetable page.
The L4 page migration is initiated 
after the corresponding data page migration is successfully 
completed (Line~\ref{algoline:dpmigrate_success}).

To migrate a \pagetable page we first fetch L4 and L3 pages 
corresponding to the new data page (\algonewdp) by performing a software \pagetable walk (Line~\ref{algoling:getmappings}).
Once we have L4 page, we get its \numa node.
We skip the migration if L4 page is already in the destination 
NUMA node (Line \ref{algoline:indest}) or if the migration is from one DRAM (or \nvmm) node
to another DRAM (or \nvmm) node (Line~\ref{algoline:indram}). 
We also skip the migration of L4 page from DRAM to \nvmm if any data page pointed 
by L4 is in DRAM (Line~\ref{algoline:hotdata}).

On meeting all the necessary conditions, we start the migration by locking L4 and L3 \pagetable pages. 
Locking is required to synchronize between parallel data or L4 migrations, which is common in multi-core systems.
Now we allocate a new L4 page ($L4_{new}$) on the destination \numa node. If successful, we flush the TLB and MMU caches to invalidate any entries pointing to old L4 page
and then copy the contents from old L4 page to $L4_{new}$ and update L3 to point to $L4_{new}$ (Line~\ref{algoline:sync-pt}).

TLB flushing forces a hardware page walk on CPUs that concurrently attempts to access the old L4 page under migration, while an invalid old L4 entry triggers a page fault. The operating system's page fault handler being aware of the ongoing L4 migration
waits for the migration to complete before inserting the updated mapping.

\subsubsection{Page table consistency}
\label{sec:consistent_pt}
In a multi-core system, multiple CPUs can concurrently try to access an L4 page under
migration in the software page fault handler.
Furthermore, similar to a data page migration, an L4 page
migration can also be triggered simultaneously, thus, requiring explicit synchronization during a
\pagetable migration.
We also need to ensure that the hardware \pagetable walker sees a consistent state of the \pagetable at all the times.

Even though Algorithm~\ref{algo:pte_migration} provides generic steps to migrate an L4 page,
the actual implementation and sequence of steps (e.g., when to flush TLB entries) may vary
depending on the underlying architecture and the operating system.

%% file: implementation.tex
\section{Implementation}
\label{sec:implementation}

In this section, we explain the implementation details of \methodname for x86\_64
architecture in the Linux kernel.
We use the Linux kernel's terminology to refer to different levels of a \pagetable; L1 is referred as PGD, L2 as PUD, L3 as PMD, and L4 as \pte. 

As explained before, the default kernel only migrates data pages during a migration. Enabling \pte migration on a
mult-core system is not trivial; a simple pointer flip at the PMD-level and freeing of the old \pte page is not enough.
We list down a few challenges in implementing \pte migrations on a production-class operating
system such as Linux:

\one Multiple CPUs in a multi-core system, upon a TLB miss, can concurrently perform page walk by accessing the \pagetable pages using the physical addresses. Hence, we need to ensure that the hardware always sees a consistent \pagetable.

\two As a \pte page points to 512 data pages, it is possible to have multiple concurrent migrations of these data pages to different \numa nodes. Every such instance of successful data migration triggers a \pte page migration. We need to ensure that the \pagetable is consistent without causing a significant performance overhead.

In the subsequent sections, we explain implementation details including challenges and solutions.

\subsection{Binding the page table pages}

The default Linux kernel allows us to specify memory policies for applications to bind to specific \numa nodes. However, Linux does not support binding \pagetable pages independent of the data pages. We modify the \pagetable page allocation functions in the kernel, \texttt{pgd\_alloc}, \texttt{pud\_alloc}, and \texttt{pmd\_alloc}, to add support to bind PGD, PUD, and PMD pages in DRAM.
 
We extend the \texttt{numactl} utility to select the 
processes for which the high-level pages of a \pagetable should be placed in DRAM.
Placement of high-level \pagetable pages is independent of data page placement for processes enabled with \texttt{numactl} binding. Rest of the processes in the system follow the data page placement policy for \pagetable pages.

\subsection{PTE migrations}
\label{sec:pte_migrations_implementation}
The Linux kernel ensures that a data page under migration is completely isolated from the rest of the system. Any page fault on this page waits either on the locked \pte or the locked data page until the migration is complete.

As shown in Figure \ref{fig:default_page_migration}, we first try to acquire the PMD lock. If successful, a new \pte page is allocated on the destination \numa node using \texttt{alloc\_pages\_node()} function. Then, we copy the page content from the old \pte page to the new \pte page and fix the \pagetable (update the PMD entry to point to this new \pte).

We also flush the TLB entries and MMU cache to clear the old PMD to PTE mappings. But, the PTE to data page mappings are still 
valid as we copy the contents of old \pte page to the new \pte page (see Figure \ref{fig:default_page_migration}).
After the PMD to new \pte page mapping is updated in the \pagetable, any TLB miss will use the new \pte page instead of the old \pte page; the hardware need not wait for the release of the lock on the old \pte page.

\subsection{Performance implications}
\label{sec:performance_pte}
The \pagetable of a process has three types of locks; a \pagetable lock, a per-PMD page lock, and a per-\pte page lock (see Figure \ref{fig:page_table}). The per-\pte page lock allows for parallel updates across different \pte pages without locking the whole \pagetable. This significantly improves the performance of operations on the last level of the \pagetable in a multi-core system~\cite{split_locks_1,split_locks_2}.

As explained in Section \ref{sec:migrating_pte_page}, we obtain the PMD lock prior to updating the PMD entries. This is required to avoid a race condition where a parallel migration on another CPU updates the PMD entry. However, locking the PMD serializes the migration of data pages mapped within the PMD with the migration of the corresponding \pte pages. This delays the completion of a data page migration, which in turn increases the page fault latency as the Linux kernel's fault handler has to wait for the completion of the migration.
To mitigate the latency overheads, we try to lock the PMD using \texttt{try\_lock()} prior to migrating a PTE page. 
  If we cannot get the lock, we skip the \pte page migration. As a \pte page points to 512 data pages, it
is possible that we will get many more opportunities to migrate the \pte page.

%% file: evaluation.tex
\newcommand{\bhigh}{BHi\xspace}
\newcommand{\ball}{BAll\xspace}
\newcommand{\bonlymig}{Mig\xspace}
\newcommand{\bmig}{BHi$+$\bonlymig\xspace}
\section{Evaluation}
\label{sec:evaluation}

In this section,  we evaluate the performance of \radiant on a suite of real-world applications and synthetic benchmarks, and compare it with the Linux kernel's memory allocation policies and management techniques. Table \ref{tab:system_config} provides details on the experiment setup. Support for transparent huge page (THP) is disabled unless otherwise stated. We use an unmodified Linux kernel 5.6 for all our baseline evaluations and enhance it to implement \radiant.
Table \ref{tab:bench} lists the workloads and 
Table~\ref{tab:conventions} lists the conventions used for the evaluation.

\begin{table}
	\centering
	\footnotesize
	\resizebox{\columnwidth}{!}{%
		\footnotesize
		\begin{tabular}{llllll}
			\multicolumn{6}{c}{\textbf{Hardware}}                                                                                                              \\ \hline
			\multicolumn{3}{|c|}{\textbf{CPUs (2$\times$24$\times$2$=$96)}}                                              & \multicolumn{3}{l|}{\textbf{Memory (2 TB)}}                             \\ \hline
			\multicolumn{1}{|l|}{Model}    & \multicolumn{2}{l|}{Intel-Xeon Gold 6252N}                 & \multicolumn{1}{l|}{DRAM}    & \multicolumn{2}{l|}{384\,GB}       \\ \hline
			\multicolumn{1}{|l|}{CPUs} & \multicolumn{2}{l|}{2 Socket, 24 Cores, 2 HT }                      & \multicolumn{1}{l|}{Optane}  & \multicolumn{2}{l|}{1.6\,TB (Flat Mode)}       \\ \hline
			\multicolumn{6}{c}{\textbf{System Setting}}                                                                                                        \\ \hline
			\multicolumn{2}{|l|}{Linux Kernel: 5.6}  & \multicolumn{2}{l|}{DVFS: Performance}                  & \multicolumn{2}{l|}{ASLR: Off}       \\ \hline
			\multicolumn{6}{c}{\textbf{NUMA: 4 Nodes}}                                                                                                         \\ \hline
			\multicolumn{3}{|c|}{\textbf{Node 0/Node 1}}                                           & \multicolumn{3}{c|}{\textbf{Node 2/Node 3}}                             \\ \hline
			\multicolumn{1}{|l|}{CPUs}     & \multicolumn{2}{l|}{48}                        & \multicolumn{1}{l|}{CPUs}    & \multicolumn{2}{l|}{0}            \\ \hline
			\multicolumn{1}{|l|}{Memory}   & \multicolumn{2}{l|}{DRAM 192\,GB}               & \multicolumn{1}{l|}{Memory}  & \multicolumn{2}{l|}{Optane 800\,GB} \\ \bottomrule
		\end{tabular}%
	}
	\caption{System configuration}
	\label{tab:system_config}
\end{table}

\begin{table}
	\centering
	\resizebox{\columnwidth}{!}{%
		\footnotesize
		\begin{tabular}{|p{1.7cm}|p{3.7cm}|p{1.3cm}|}
			\hline
			\textbf{Name} & \textbf{Description}&   \textbf{RSS Size (Page table size)}  \\ \hline
			Memcached~\cite{memcached}     & A commercial distributed in-memory object caching system. \tiny{Setting: YCSB~\cite{ycsb}: 2\,M objects. Read using a Zipfian distribution ~\cite{zipfian}.}                       &  1\,TB (1.9\,GB) \\ \hline
			Redis~\cite{redis}         & A commercial in-memory key-value store.\newline \tiny{Setting: Same as \memcached.}                                            &  1\,TB (1.9\,GB)         \\ \hline
			BTree~\cite{btree}         & A benchmarks for index look-ups used in database and other large applications.
			\newline \tiny{Setting: 7.3\,B elements with 40\,M look-ups.}         &   666\,GB (1.2\,GB)     \\ \hline
			HashJoin~\cite{hashjoin} & A benchmark for hash-table probing used in database applications and other large applications.
			\newline \tiny{Setting: 6\,B elements.}   & 838\,GB (1.6\,GB)  \\ \hline
			XSBench~\cite{xsbench} &  A key computational kernel of the Monte Carlo neutron transport algorithm~\cite{xsbench} \newline \tiny{Setting: 2\,M grid points.}   & 1\,TB (1.9\,GB)\\ \hline
			BFS~\cite{bfs}          &  A graph traversal algorithm. \newline \tiny{Setting: rMat order 30 graph~\cite{bfs}}                                                     &   600\,GB (1.1\,GB)  \\ \hline
			
		\end{tabular}%
	}
	\caption{Workloads used to evaluate the performance of \methodname}
	\label{tab:bench}
\end{table}

\begin{table}
	\centering
	\resizebox{\columnwidth}{!}{%
		\footnotesize
		\begin{tabular}{|l|p{6.5cm}|}
			\hline
			\multicolumn{2}{|c|}{\textbf{\radiant techniques}} \\ \hline
			\bhigh              & Bind high-level (PGD, PUD and PMD) \pagetable pages in DRAM  \\ \hline
 			\bonlymig              & Enable migration of last-level (PTE) \pagetable pages \\ \hline
			\bmig & Enabling binding of high-level \pagetable pages along with migration for the last-level of the \pagetable.  \\ \hline
		\end{tabular}%
	}
	\caption{Conventions used in the paper for discussion}
	\label{tab:conventions}
\end{table}

\subsection{Evaluation strategy}

We contrast the performance of \radiant techniques with two memory allocation policies in the default Linux kernel. 
\one First is the default first-touch policy ~\cite{numa_challenges,mitosis}. In this case, the \numa node for the \pagetable pages are selected based on the data page allocation policy, i.e., a \pagetable page is allocated on the same \numa node where the data page is allocated. This policy allocates a data page close to the CPU that is running the application; however, allocations can spill over to remote nodes when an allocation request cannot be severed from the local \numa node~\cite{numa_challenges}. 

\two Second is the interleaved policy where the Linux kernel distributes the data uniformly across all the \numa nodes in a round-robin order to improve memory bandwidth utilization.

To enable PTE migrations, we rely on the Linux kernel's memory management technique called \autonuma to get data page migration hints. By default, \autonuma dynamically migrates data pages only (not page table pages) across \numa nodes to improve local \numa accesses from a CPU. We run the experiments with \autonuma enabled unless otherwise mentioned. \\

\noindent
Our evaluation strategy is as follows:

\begin{itemize}
	\item \textbf{Full-system run: } Run the workloads with full system capacity utilizing maximum possible resources, which reflects a typical real-world data center scenario. We compare the performance of \radiant (\bhigh and \bmig) with Linux kernel's first-touch policy, and show that enabling \pte migrations improves the performance.
	
	\item \textbf{Multi-tenant scenario:} Evaluate the performance benefits of \radiant in a multi-tenant environment (a typical cloud setting), where different applications can start and exit at any point in time. We show that enabling PTE migration (\bmig) for applications significantly improves performance.
	
	\item \textbf{Interleaved setting:} Compare the performance of \radiant (\bhigh) with the interleaved memory allocation policy, with \noautonuma. We show that differentiating between allocation of data and \pagetable pages improves the performance.

	\item \textbf{Start up time:}  At the startup of a large memory footprint application, a significant portion of high-level (PGD, PUD, and PMD) \pagetable pages are initialized. We evaluate the performance benefits of \bhigh in such scenarios.

	\item \textbf{Huge page impact:}  Evaluate the performance benefits of \radiant when huge pages are enabled.
\end{itemize}

\begin{figure*}
	\begin{subfigure}[t]{.34\linewidth}
		\raggedleft
		\includegraphics[width=1.02\linewidth]{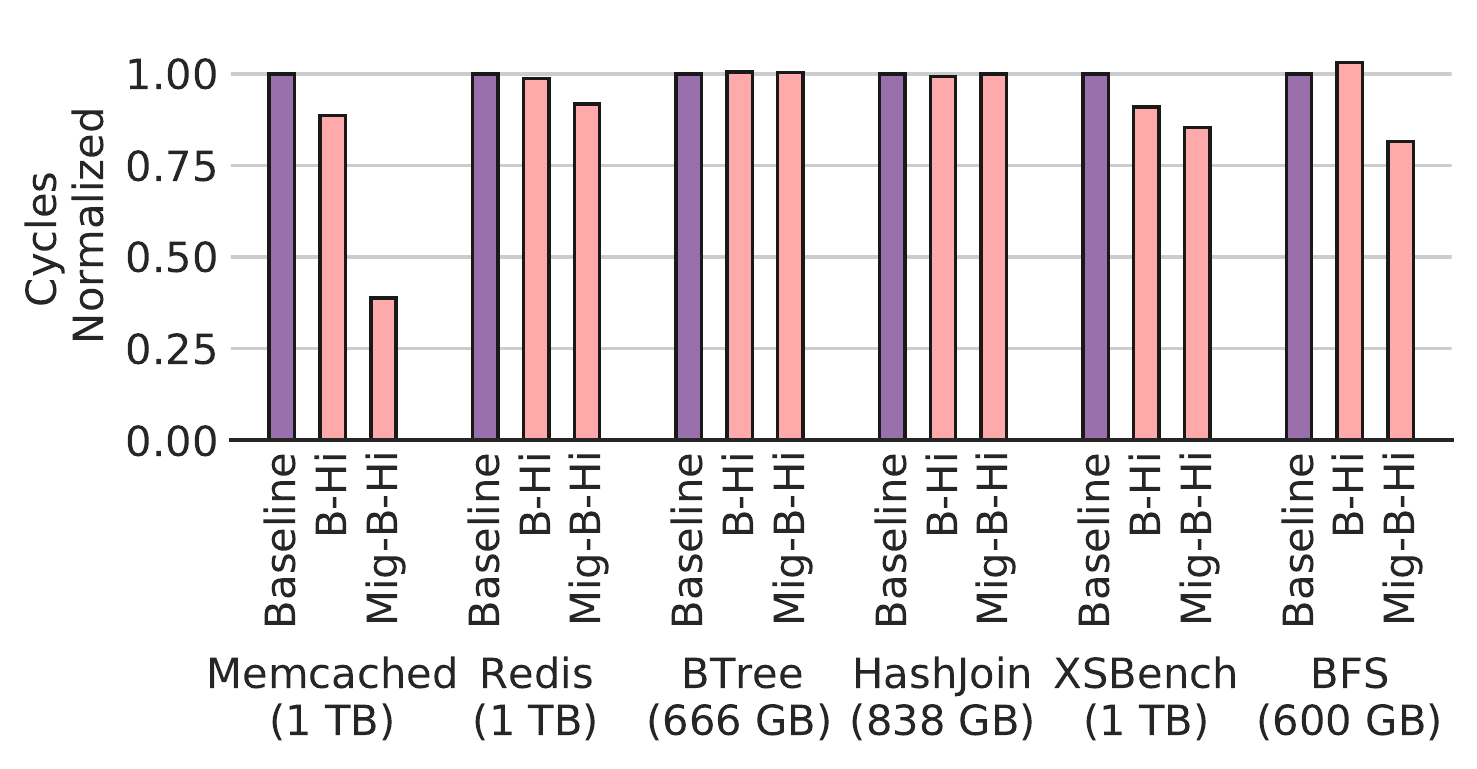}
		\caption{Total cycles}
		\label{fig:compare_an_main}
	\end{subfigure}
	\begin{subfigure}[t]{.32\linewidth}
	\includegraphics[width=\linewidth]{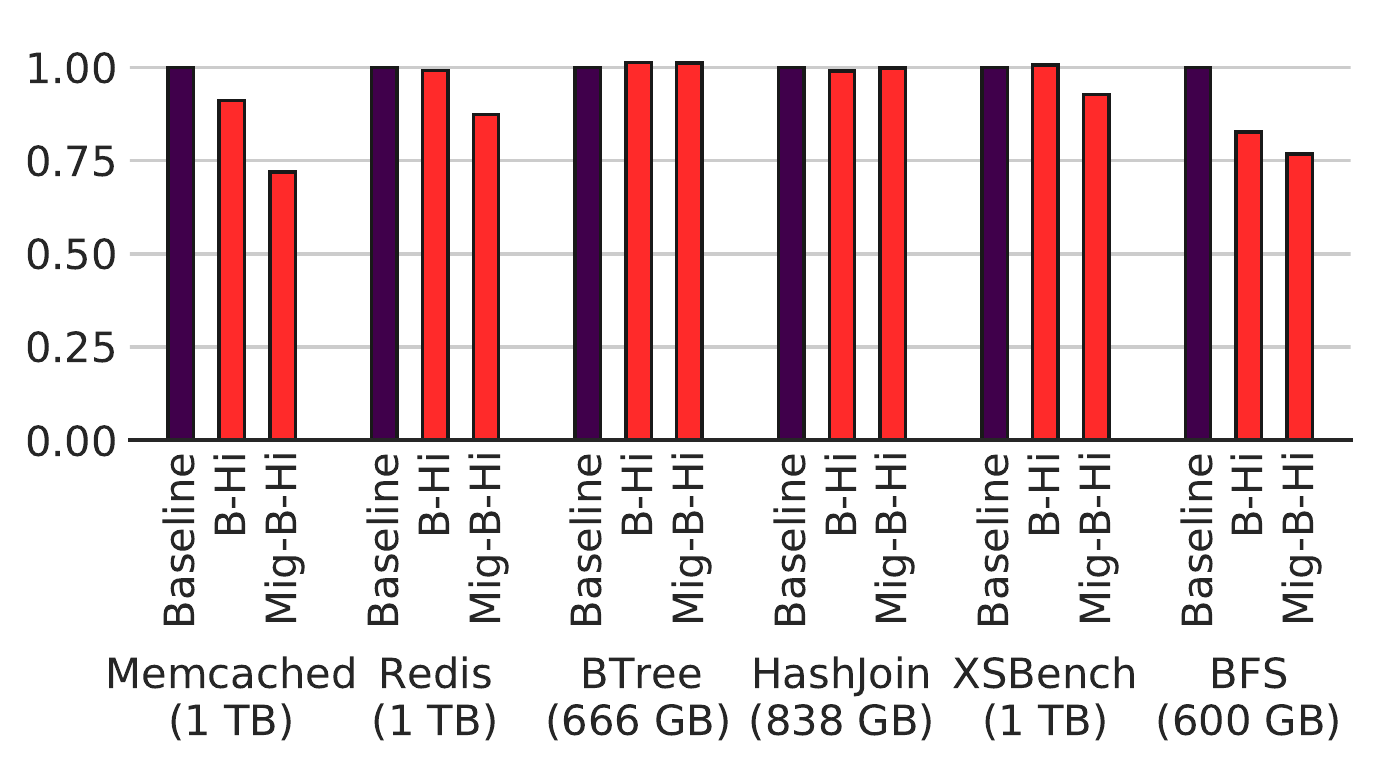}
	\caption{Walk cycles }
	\label{fig:compare_an_walk}
\end{subfigure}
	\begin{subfigure}[t]{.32\linewidth}
	\includegraphics[width=\linewidth]{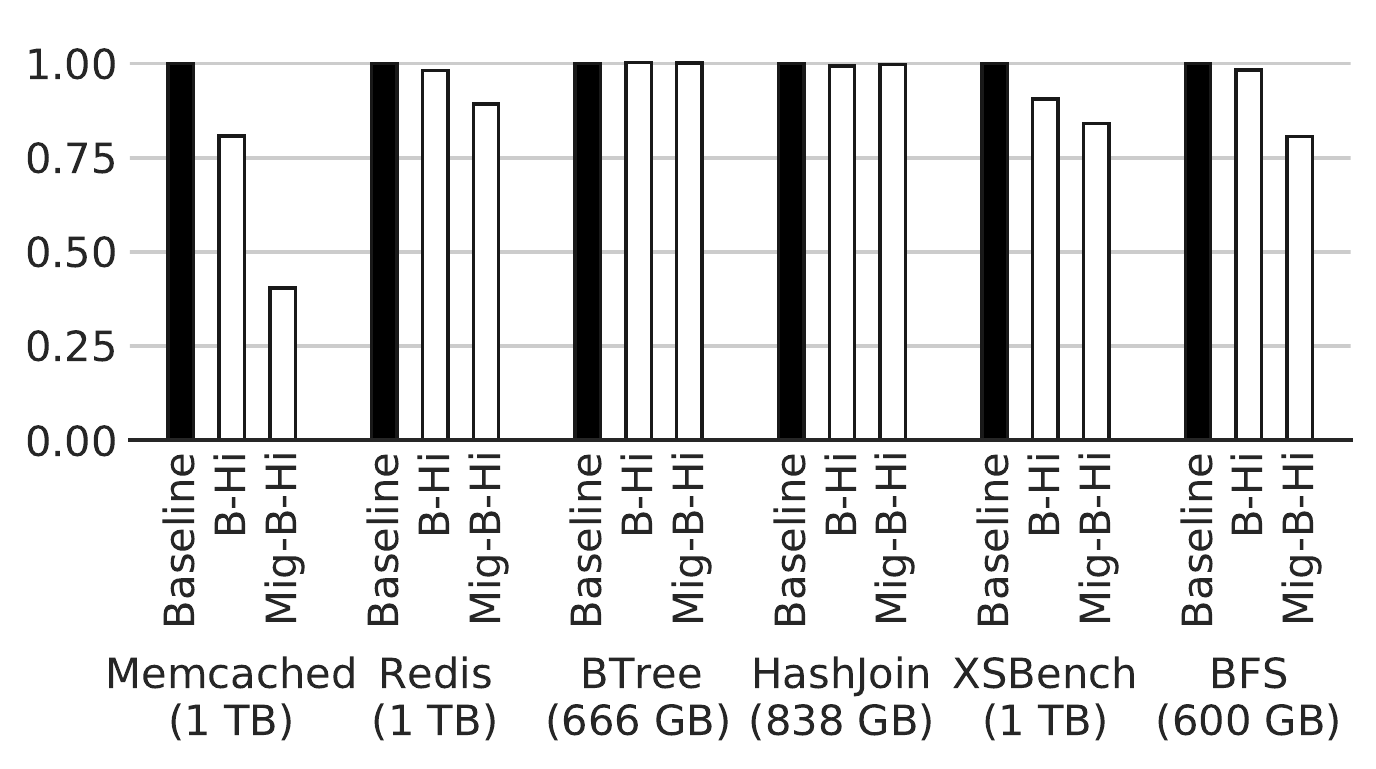}
	\caption{Stall cycles }
	\label{fig:compare_an_stall}
\end{subfigure}

	\caption{Performance comparison of first-touch policy with \radiant, for the run-phase. Bars are color-coded to differentiate between the Linux kernel's and \radiant techniques.}
	\label{fig:compare_an}
\end{figure*}

\begin{table}[!ht]
	\centering
	\resizebox{\columnwidth}{!}{%
		\footnotesize
		\begin{tabular}{lrrrr}
			\toprule
			& \textbf{Run Time} & \textbf{Cycles} & \textbf{Walk Cycles} & \textbf{Stall Cycles} \\
			\midrule
			\multicolumn{5}{c}{Full system run:  First-touch policy} \\ \hline
			\textbf{\bhigh } &    2.79\% &  3.32\% &       4.56\% &        5.68\% \\
			\textbf{\bmig} &    20.39\% & 20.71\% &      12.38\% &        20.9\% \\ 
			\hline \hline 
			\multicolumn{5}{c}{Multi-tenant scenario:  First-touch policy} \\ \hline
			\textbf{\bmig} &   17.95\% & 19.85\% &      32.62\% &       23.25\% \\ 
			\hline \hline 
			\multicolumn{5}{c}{Interleaved: \noautonuma, Interleaved policy} \\ \hline
			\textbf{\bhigh} &    10.41\% & 10.02\% &      10.53\% &        9.01\% \\
			\hline \hline 

			\multicolumn{5}{c}{Huge page impact: \noautonuma with THP enabled} \\  \hline 
			\textbf{\bhigh} & 52.96\% & 51.82\% &      36.37\% &       38.63\% \\ \hline \hline
			\multicolumn{5}{c}{Start up time improvement: \noautonuma (\redis)} \\  \hline 
		\end{tabular}

	}
\resizebox{\columnwidth}{!}{%
\footnotesize
	\begin{tabular}{rrrrrrr}
			&\textbf{Time}  &  \textbf{Avg Lat.} & \textbf{ Max Lat.} &   \textbf{$\mathbf{95^{th}\%}$ile Lat.}&   \textbf{$\mathbf{99^{th}\%}$ile Lat.} \\
			\midrule
			\textbf{\bhigh}&22.81\% &    22.82\% &  17.35\% &  25.56\% &  20.70\% \\
			\bottomrule
		\end{tabular}
}
	\caption{\radiant performance improvement summary (geometric-mean across all the workloads). A higher value indicates better performance improvement with \radiant.}
	\label{tab:compare_an}
\end{table}

\begin{figure*}[!b]
	\begin{subfigure}[t]{.34\linewidth}
		\raggedleft
		\includegraphics[width=1.02\linewidth]{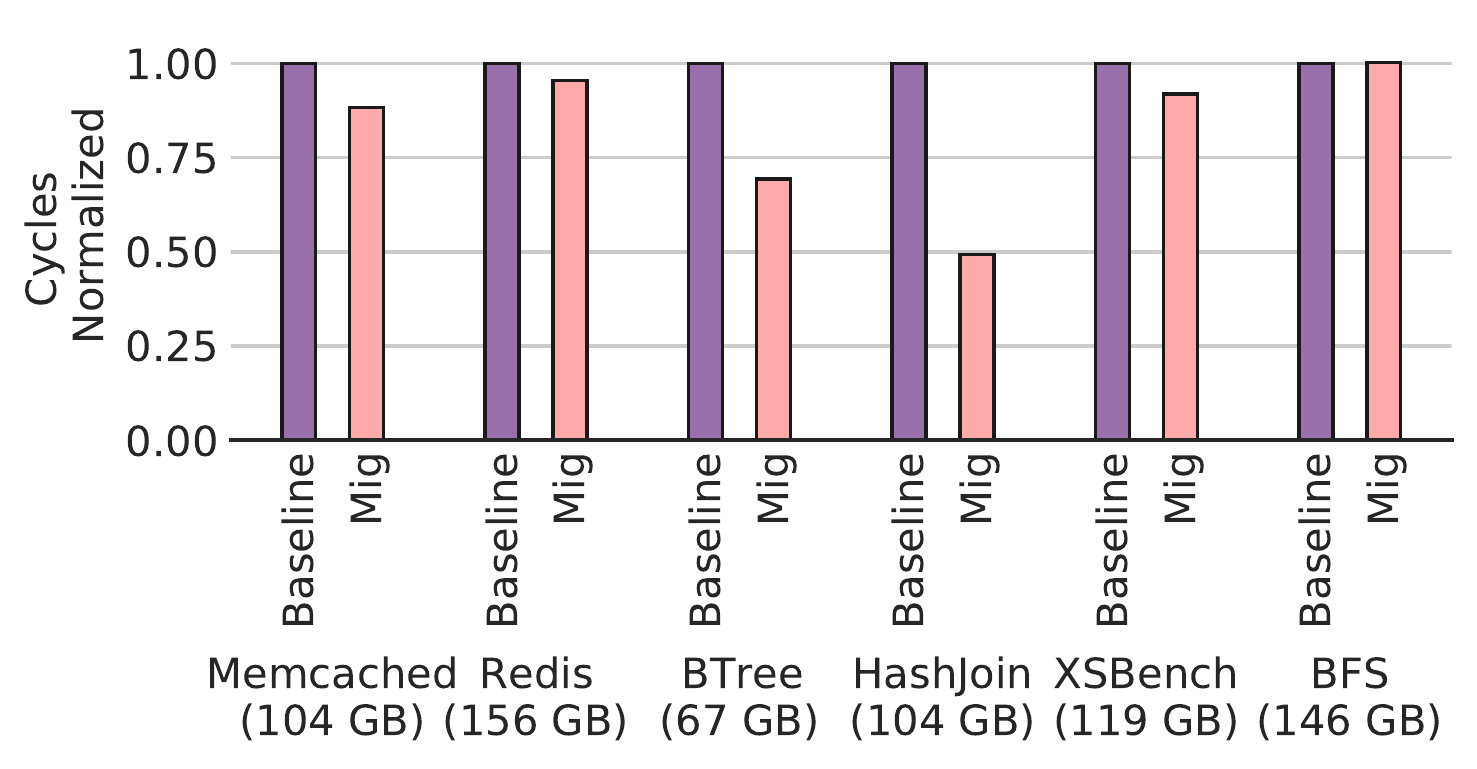}
		\caption{ Total cycles}
		\label{fig:compare_an_main_controlled}
	\end{subfigure}
	\begin{subfigure}[t]{.32\linewidth}
	\includegraphics[width=\linewidth]{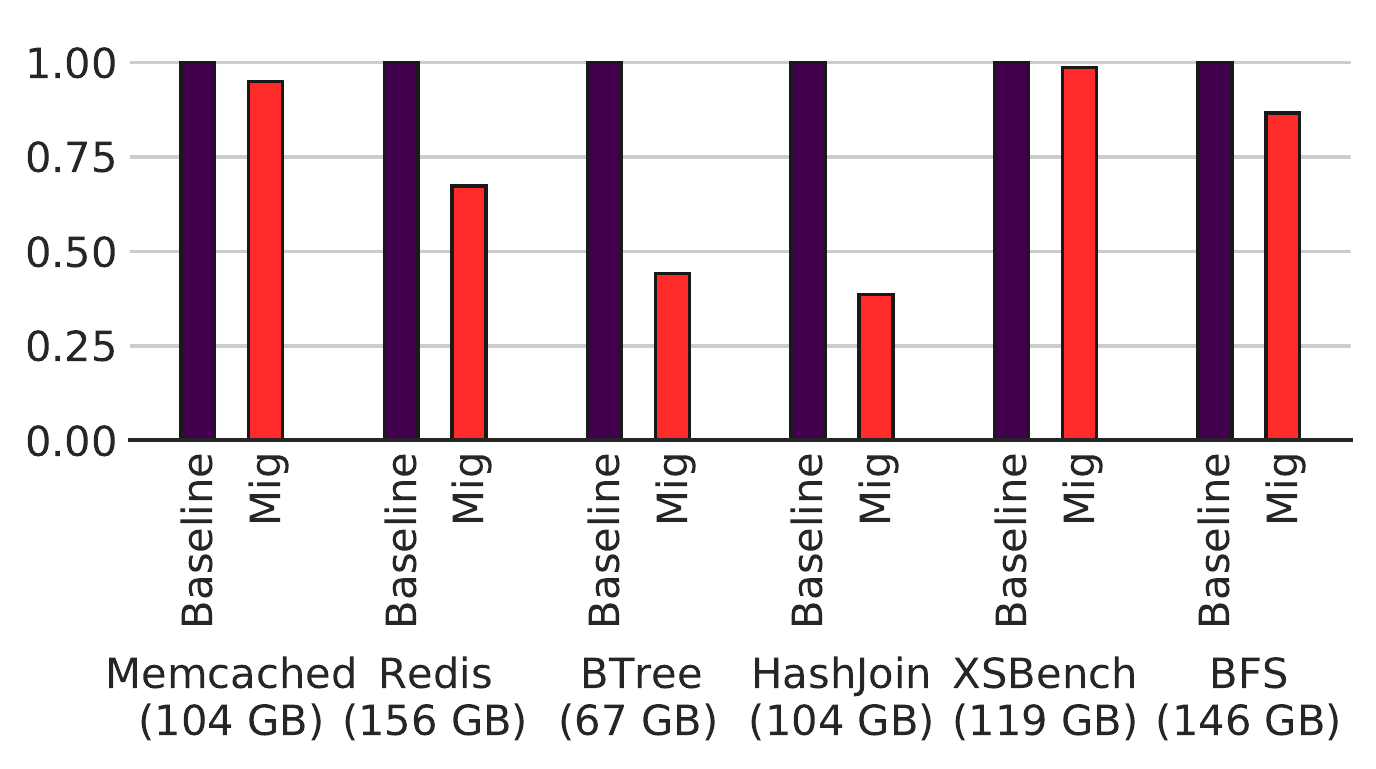}
	\caption{Walk cycles}
	\label{fig:compare_an_walk_controlled}
\end{subfigure}
	\begin{subfigure}[t]{.32\linewidth}
	\includegraphics[width=\linewidth]{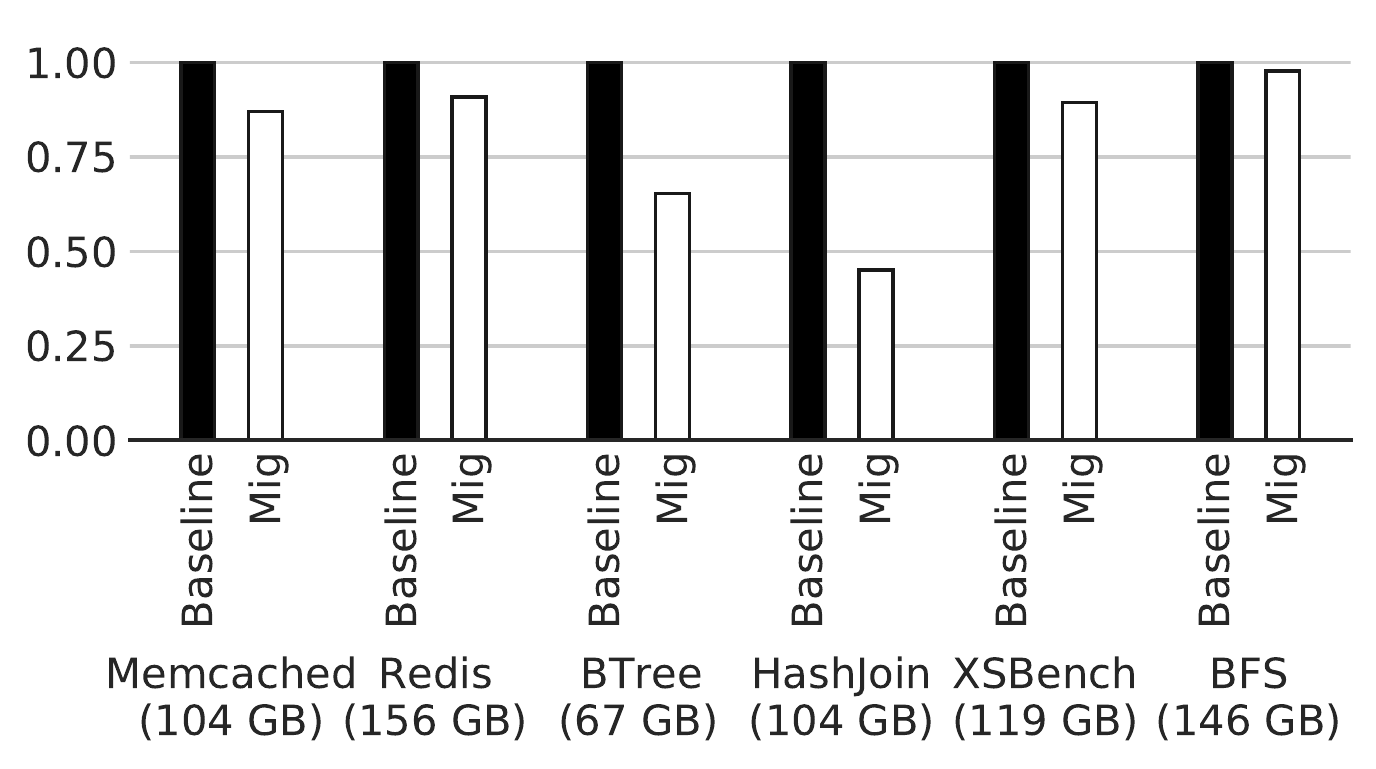}
	\caption{Stall cycles}
	\label{fig:compare_an_stall_controlled}
\end{subfigure}

	\caption{Performance comparison of \radiant (\bonlymig) in a multi-tenant environment with \autonuma (baseline). Bars in (\subref{fig:compare_an_main_controlled}) show total cycles. Bars are color-coded to differentiate between the Linux kernel's and \radiant techniques.}
	\label{fig:controlled_exp_result}

\end{figure*}

\subsection{Full-system run}
\label{sec:eval_exp1}
We evaluate the performance with the memory footprint size as specified in Table \ref{tab:bench} utilizing maximum possible system resources.
We compare the performance of the Linux kernel's first-touch policy (baseline) with \methodname (\bhigh and \bmig~) techniques (see Figure~\ref{fig:compare_an}).

\leavevmode
\newline
\noindent
\textbf{\bhigh :} The high-level \pagetable pages are frequently accessed during a \pagetable walk. Binding them to DRAM ensures a low-latency access during a \pagetable walk and reduces the walk cycles by up to 17.31\%. Placement on DRAM also reduces the stall cycles by up to 19.18\%. This translates into a reduction of total cycles by up to  11.43\%	and a  run-time improvement of up to 9.08\% (see Table \ref{tab:compare_an}). 

\leavevmode
\newline

\noindent	
\textbf{\bmig :} With \pte migrations enabled, the percentage of \pagetable pages in DRAM increases (e.g., from 19.6\% to 34.0\% for \redis). This reduces the walk cycles by up to  28.06\% and the stall cycles by up to 59.57\%. This causes a reduction in the total cycles by up to 61.19\% and improves the run-time by up to 60.88\% (see Figure \ref{fig:compare_an_main}). 

\subsection{Multi-tenant scenario}
\label{sec:eval_exp4}

In a typical cloud setting, where tiered memory is likely to be deployed, many applications co-exits in parallel in a given period of time. Here, different applications may start or exit at any time.

An application (\textit{V}) started when DRAM is almost full is allocated memory (data and \pagetable pages) on \nvmm. At a later point in time when other applications using DRAM exit, DRAM becomes free resulting in the  migration of the data pages of \textit{V} from \nvmm to DRAM. However, with the default Linux kernel, page table pages are not migrated from \nvmm, incurring performance overheads even in spite of free memory in DRAM.
To capture the benefits of \radiant in such scenarios, we setup a cloud-like environment and contrast the performance of \methodname with the default Linux kernel.

To setup the environment, we first launch applications that fills up DRAM. These  applications also frequently access the data pages in DRAM. Then we launch our benchmark application. As DRAM memory is full, all the benchmark application's memory is allocated on \nvmm.
After this, we terminate the applications that filled up DRAM resulting in freeing of significant portion of DRAM memory. This triggers migration of the benchmark application's data pages from \nvmm to DRAM.

\begin{table}[!ht]
	\centering
	\resizebox{\columnwidth}{!}{%
		\begin{tabular}{lR{1.6cm}|R{2cm}R{1.7cm}R{2cm}}
			\hline
			\multicolumn{2}{c|}{} & \multicolumn{3}{c}{\textbf{PTE migrations}} \\ \hline
			\multicolumn{1}{l}{Workload} & Data page migrations & Successful migration & Already in destination & With in DRAM \\
			\hline
			Memcached         & 66,644,738        & 50,601                & 39,272,431         & 26,763,450     \\
			Redis             & 33,315,590        & 69,731                & 27,461,927         & 5,783,941      \\
			BTree             & 11,820,636        & 17,061                & 7,791,351          & 4,012,020      \\
			HashJoin          & 1,945,151         & 50,209                & 1,867,027          & 27,915        \\
			XSBench           & 371,977          & 574                  & 285,933           & 85,470        \\
			BFS               & 6,967,564         & 20,957                & 6,942,269          & 4,338         \\ \hline
		\end{tabular}%
	}
	\caption{Number of data page and \pte migrations in the controlled setting.}
	\label{tab:migrate_stats_controlled}
\end{table}

For this experiment, the system configurations remain the same as full-system run, however we run with a smaller input size (see Figure~\ref{fig:controlled_exp_result}). 
\bmig reduces the walk cycles by up to 61.34\% and stall cycles by up to 54.88\%. This reduces the total cycles by up to 50.75\% and improves the run-time by up to 50.77\% (see Figure~\ref{fig:controlled_exp_result}). Table~\ref{tab:migrate_stats_controlled} shows the number of data page migrations triggered and the number of successful \pte migrations. We also show the reason for not migrating a \pte page (a \pte page is already in DRAM or in the destination \numa node). As a \pte page points to 512 data pages, the first data page that is migrated to DRAM triggers
a \pte page migration; for the rest 511 data page migrations, \pte migration is not required as it is already in DRAM.

\subsection{Interleaved vs. \radiant}
\label{sec:eval_exp2}

\begin{figure}
	\begin{subfigure}[t]{.34\linewidth}
		\raggedleft
		\includegraphics[width=.9\linewidth]{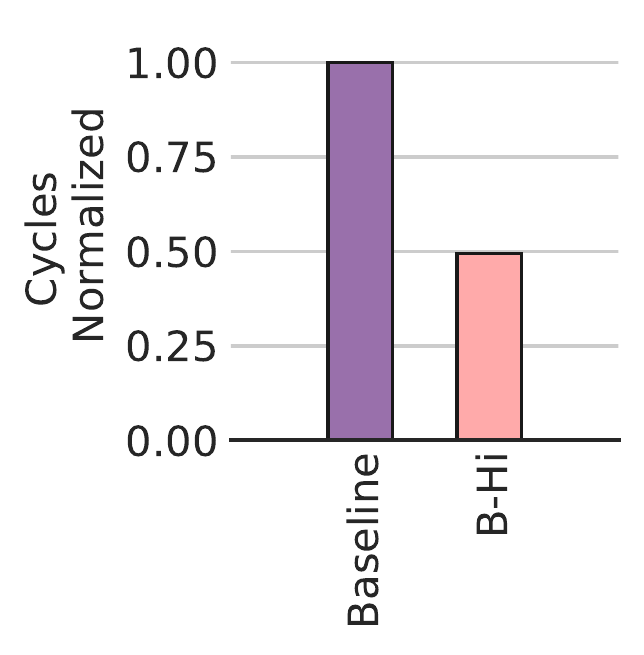}
		\caption{ Total cycles}
		\label{fig:exp2_1_pltidx_0_2}
	\end{subfigure}
	\begin{subfigure}[t]{.32\linewidth}
	\includegraphics[width=.81\linewidth]{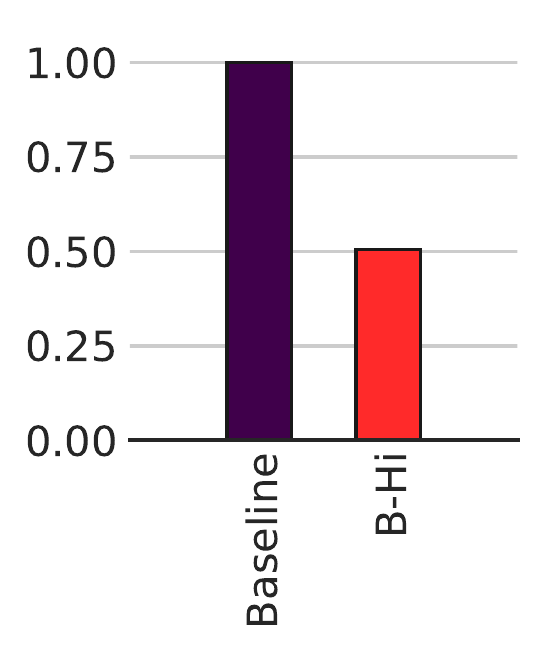}
	\caption{Walk cycles}
	\label{fig:exp2_1_pltidx_1_2}
\end{subfigure}
	\begin{subfigure}[t]{.32\linewidth}
	\includegraphics[width=.81\linewidth]{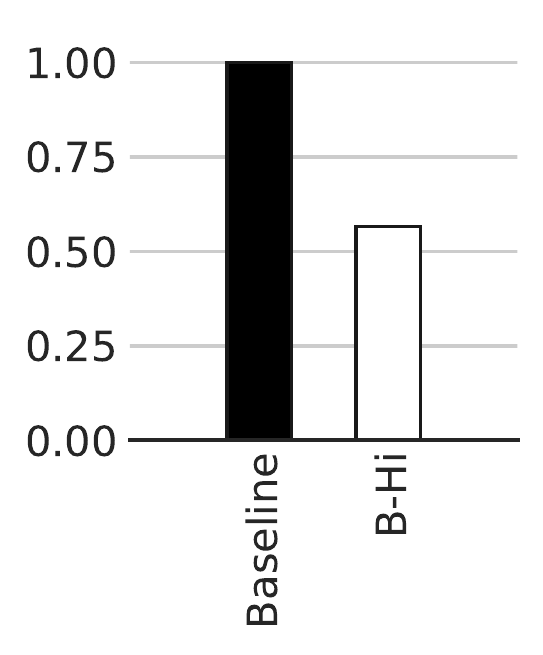}
	\caption{Stalls}
	\label{fig:exp2_1_pltidx_5_2}
\end{subfigure}

	\caption{Performance evaluation of \radiant (\bhigh) for \memcached in an interleaved setting with \noautonuma.}
	\label{fig:exp2_result}

\end{figure}

Interleaved memory allocation policy allocates the \pagetable pages and the data pages on DRAM and \nvmm in a round robin manner. 
\radiant still follows the interleave policy for data pages, but binds the high-level  \pagetable pages to DRAM. 
We compare the performance of \bhigh with the default kernel allocation (Figure~\ref{fig:exp2_result}). 
As \autonuma is disabled for this experiment, \pagetable pages are not migrated and hence we do not report \bmig statistics. 
We can clearly observe that having a different placement and allocation policy for data and \pagetable pages is beneficial.

\begin{figure}[!ht]
	\centering
	\includegraphics[width=.8\linewidth]{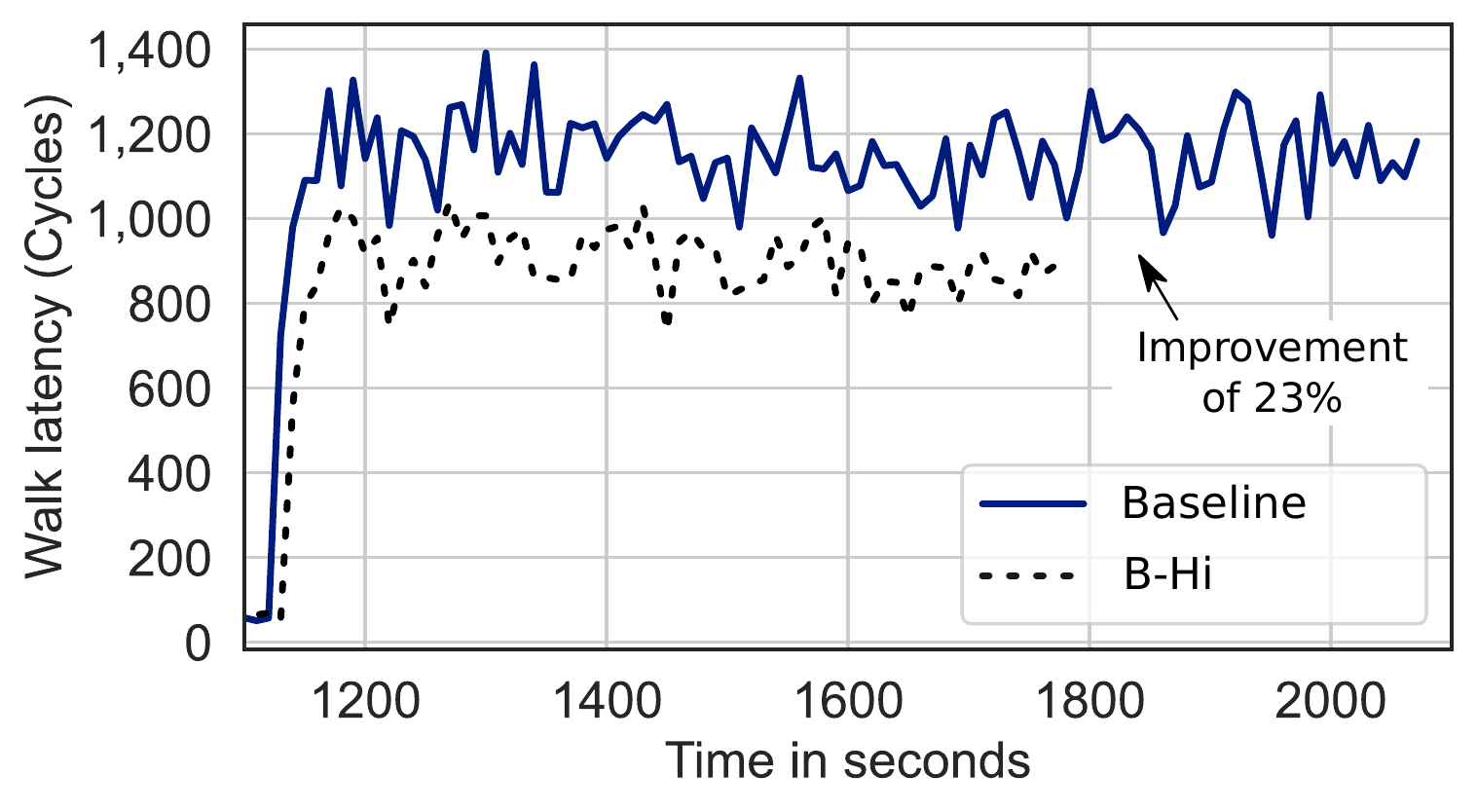}
	\caption{Improvement in the page walk latency with \bhigh for the interleaved policy (\memcached, RSS 1\.TB, 100\% read). Baseline is \noautonuma with interleaved memory allocation policy. }
	\label{fig:redis_latency_ops}
\end{figure}

\leavevmode
\newline
\noindent
\textbf{\bhigh :} Binding the high-level pages in DRAM reduces the walk cycles up to 49.48\% and stall cycles by up to 43.42\%. This reduces the total cycles by up to 50.51\% and improves the run-time by up to 51.75\%. It can be further observed from
Figure~\ref{fig:redis_latency_ops} that page walk latency decreases when we bind the high-level \pagetable
pages in DRAM as the interleaved allocation policy spreads the high-level \pagetable across the DRAM and \nvmm nodes.

\subsection{Improving application start up time}
\label{sec:eval_exp3}
During an application start up there are many data page faults that requires a page table walk. By placing the high-level of page table pages in DRAM, we reduce the cycles spent on page table walks.
While inserting 1\,TB of data in Redis, we reduce the total page walk cycles by $\approx 9\%$. This results in a $21\%$ reduction in total stalls cycles, that corresponds to an improvement of $22\%$ in total start up time, when compared with default first-touch policy (see Figure \ref{fig:default_binded_ops} and Table~\ref{tab:compare_an}).

\subsection{Huge page impact}
\label{sec:eval_hugepages}
\label{sec:eval_exp5}
We evaluate the performance of \radiant when transparent huge page (THP) support is enabled.

\begin{figure}
	\begin{subfigure}[t]{.34\linewidth}
		\raggedleft
		\includegraphics[width=.9\linewidth]{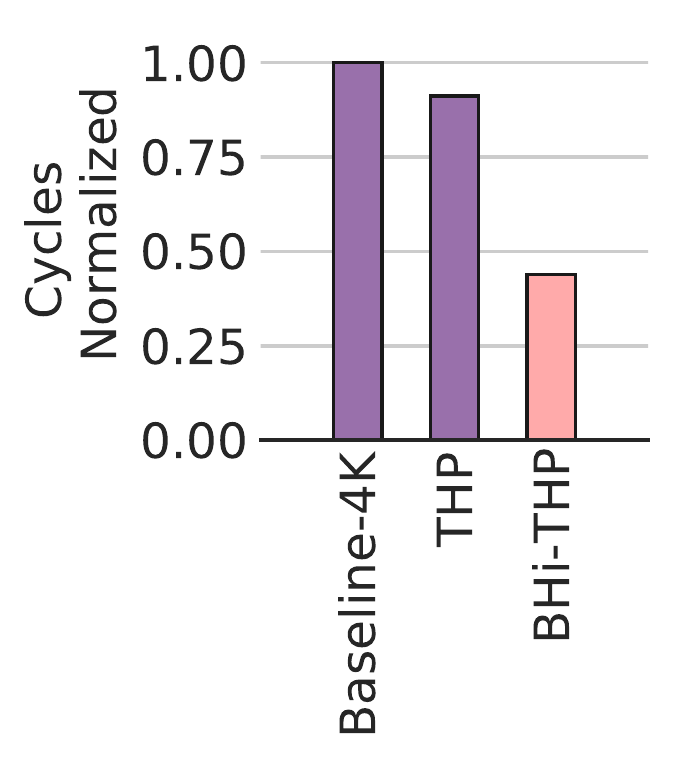}
		\caption{ Total cycles}
		\label{fig:exp1_hp_1_1_pltidx_0_2}
	\end{subfigure}
	\begin{subfigure}[t]{.32\linewidth}
	\includegraphics[width=.81\linewidth]{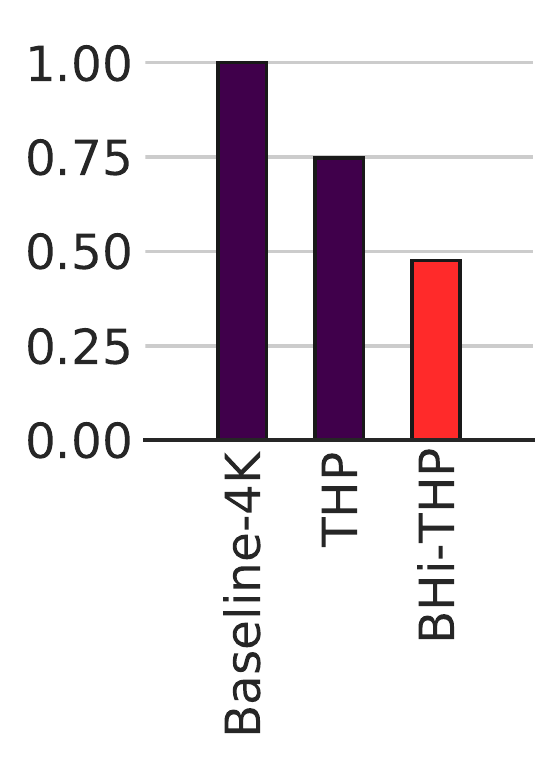}
	\caption{Walk cycles}
	\label{fig:exp1_hp_1_1_pltidx_1_2}
\end{subfigure}
	\begin{subfigure}[t]{.32\linewidth}
	\includegraphics[width=.81\linewidth]{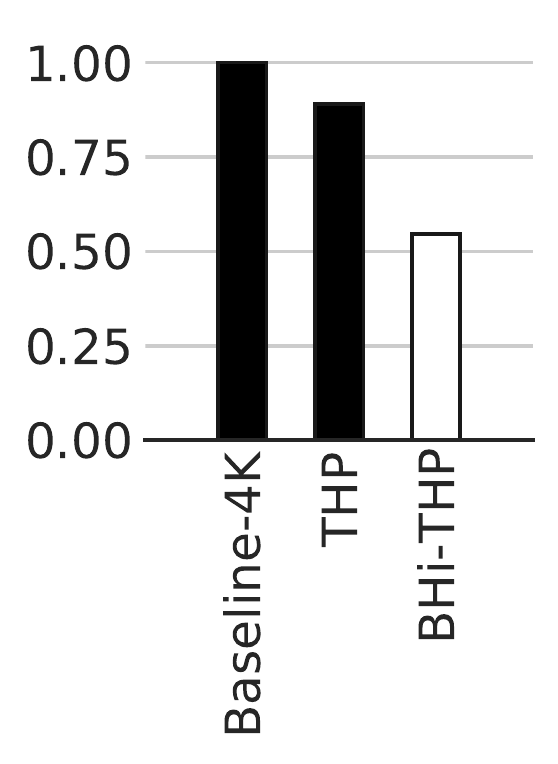}
	\caption{Stalls}
	\label{fig:exp1_hp_1_1_pltidx_5_2}
\end{subfigure}

	\caption{Performance compare to \noautonuma using base 4\,K pages (baseline).}
	\label{fig:huge_page_1}

\end{figure}

Figure~\ref{fig:huge_page_1} shows that \bhigh improves performance when THP is enabled.
\bhigh binds PGD, PUD, and PMD levels of the page table to DRAM. For a huge page as a PMD page is the last or leaf-level page (no \pte page), \bhigh is effectively binding the entire \pagetable resulting in performance improvement.
However, \bmig does not improve performance as there are no \pte-level pages to migrate.

\subsection{Discussions}
\label{sec:discussions}	

In a modern out-of-order CPU, a \pagetable walk performed by
the Page Miss Handler (PMH) in the hardware can overlap with other work~\cite{mitosis}.
Hence, a reduction in \pagetable walk cycles need not always result in the
reduction in execution cycles. We use the
hardware performance counters to reason and understand the
impact of reduction in \pagetable walk cycles.

\begin{figure}[!ht]
	\centering
	\includegraphics[width=.8\linewidth]{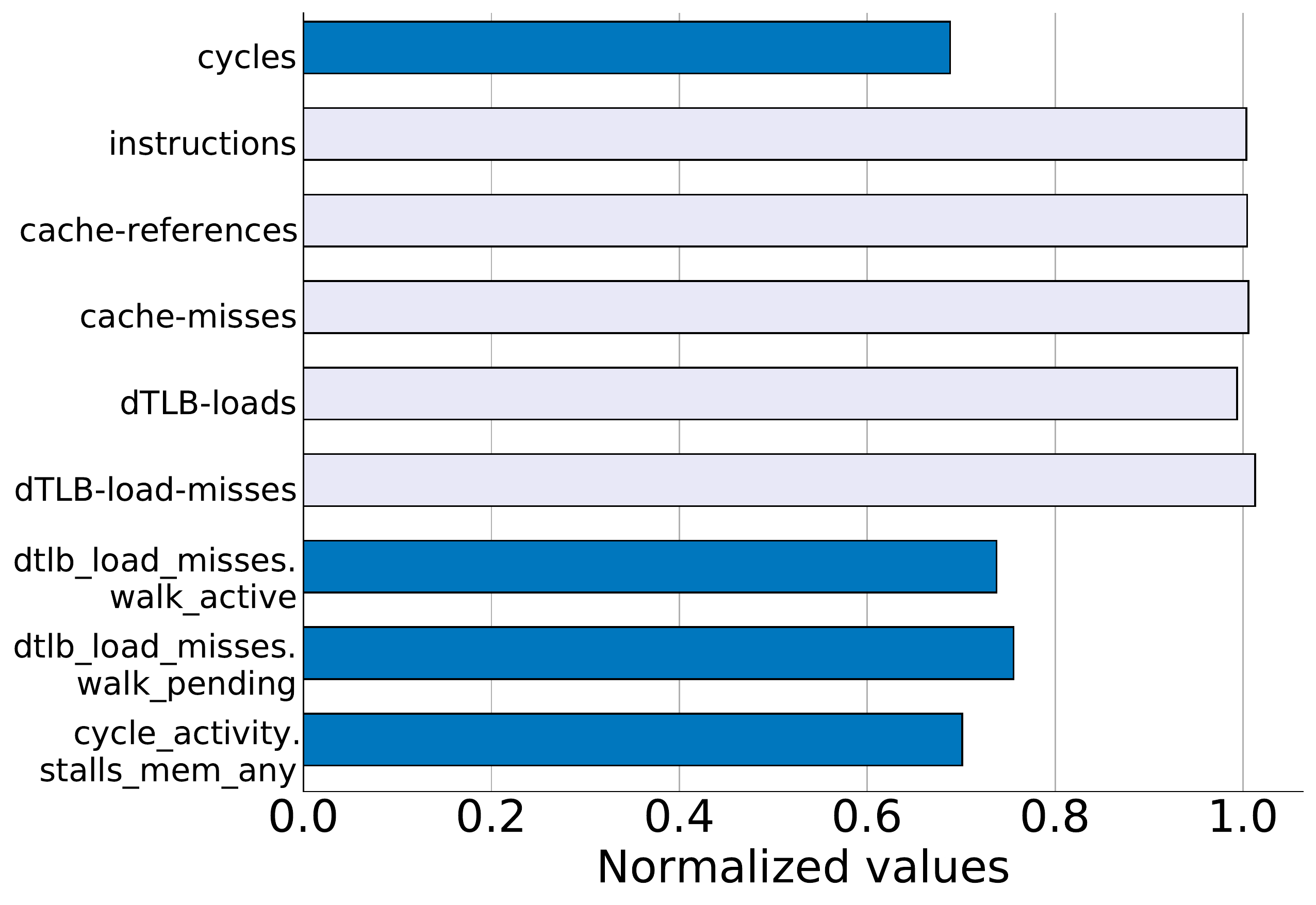}
	\caption{Performance statistics from the \texttt{perf} tool for \bmig in Experiment 1 for \bfs (normalized to default first-touch policy baseline).}
	\label{fig:perf_events}
\end{figure}

Figure~\ref{fig:perf_events} shows the counters for BFS from full system run (\S \ref{sec:eval_exp1}).
It can be observed that the instructions executed,
cache-misses incurred, and data TLB loads/load-misses remain the same as expected.
However, we can observe a significant reduction in
\texttt{walk\_active} and \texttt{walk\_pending} cycles (i.e., cycles when PMH is
busy with a page walk). This significantly contributes to 
the reduction in \texttt{stalls\_mem\_any} (execution stalls 
due to an outstanding load), which in turn reduces the total execution cycles.

We observe that for few benchmarks, a reduction in the walk cycles does not results
in a proportional reduction in the stall cycles
(\xsbench in Figure~\ref{fig:compare_an_stall}, \redis and \btree in Figure~\ref{fig:compare_an_stall_controlled}).
As a result, we do not see significant reduction in the total execution cycles.

%% file: relatedwork.tex
\section{Related works}
\label{sec:relatedworks}
\subsection{Mitosis}
Mitosis~\cite{mitosis} proposes to reduce the \pagetable overheads in a multi-socket NUMA systems 
by transparently replicating the \pagetable pages on all the NUMA nodes. Mitosis shows 
that accessing \pagetable pages from a remote NUMA node increases the page-fault latency.
The basic assumption is that all sockets are equipped with low-latency DRAM memory. However, in a tiered-memory system with high latency
NVMMs, replicating \pagetable pages has several disadvantages. First, replicating a
\pagetable and ensuring its consistency on NVMMs incurs high overheads. Second, accesses to a \pagetable on
local NVMM-backed NUMA nodes are costly due to 3$\times$ higher access latency.
Hence, replication of \pagetable may not be helpful for large memory footprint 
applications running on large capacity tiered memory systems.

Even though Mitosis supports migration of
\pagetable pages, it is achieved via replication, i.e., replicate the \pagetable
on the destination node and then lazily free the replica on the local node.
\methodname binds critical parts of the \pagetable in DRAM and dynamically migrates
the L4 pages pages between DRAM and NVMM; thus avoiding a full \pagetable migration
(Table~\ref{tab:compare_mitosis}).

Finally, \methodname employs the novel data-page-migration triggered \pagetable page migration technique to identify and migrate \pagetable pages between DRAM and NVMM. Mitosis
neither identifies nor migrates relevant \pagetable pages.

\begin{table}[!ht]
	\centering
	\resizebox{\columnwidth}{!}{%
		\footnotesize
		\begin{tabular}{ccc}
			\hline
			& {\bf \methodname}    & {\bf Mitosis}         \\ \hline
			\hline
			Tiered Memory Support       & Yes            & No              \\ \hline
			Migration Support           & Direct         & Via replication  \\ \hline
			Migration b/w DRAM and NVMM  & Yes         & No \\ \hline
			Migration Granularity       & L4 pages only  & Full \pagetable \\ \hline
			Page table DRAM binding     & L1, L2, L3     & None            \\ \hline
			Replication                 & No             & Yes             \\ \hline
			Page table sync. overheads   & No             & Yes             \\ \hline
			Hot PTE page identification & Yes            & No              \\ \hline
		\end{tabular}%
	}
	\caption{Comparison of \methodname with Mitosis}
	\label{tab:compare_mitosis}
\end{table}

\subsection{Linux kernel community}

Linux kernel patches~\cite{linux_pgtable} posted in the Linux Kernel Mailing List (LKML)
propose to bind all the \pagetable pages in DRAM to avoid accessing it from NVMM
(this patch is not a part of the Linux kernel). However, such an
approach results in pathological behaviour mentioned in \S\ref{motivation:binding}. 
\methodname proposes to bind only 0.18\% of the \pagetable pages in DRAM (i.e., L1, L2 and L3 pages)
and dynamically migrates L4 pages between DRAM and NVMM.

%% file: conclusion.tex
\section{Conclusion}
\label{sec:conclusion}
In this paper, we show that explicit and efficient management of \pagetable
on tiered memory systems with terabytes of memory is important. 
We study the performance impact of \pagetable
placement and argue that different placement and migration 
policies are required for data and \pagetable pages. 
We demonstrate that binding a small but
critical \pagetable pages to DRAM and dynamically managing the rest of the 
\pagetable pages by enabling migration
results in significant performance improvement
on systems with terabytes of \nvmm memory.